\documentclass[10pt,a4paper]{article}
\usepackage[hmarginratio=1:1,vmarginratio=1:1,margin=2cm,top=2cm]{geometry}
\usepackage[utf8]{inputenc}
\usepackage[T1]{fontenc}
\usepackage{textcomp}
\usepackage{graphicx}
\usepackage{subcaption}
\usepackage{amssymb}
\usepackage{float}
\usepackage{multirow}

\usepackage[numbers]{natbib}
\usepackage{caption}

\usepackage{amsmath}
\usepackage{makecell}
\usepackage{textcomp}
\usepackage{setspace}

\usepackage{lineno}
\usepackage{array}
\usepackage{hyperref}
\usepackage{orcidlink}
\usepackage{authblk}
\usepackage{tabto}
\NumTabs{8}
\usepackage{xcolor}

\usepackage{tikz}
\usepackage{everypage}
\usepackage{xcolor}

\AddEverypageHook{%
  \begin{tikzpicture}[remember picture, overlay]
    \fill[left color=gray!5,right color=gray!10] 
      ([yshift=0cm]current page.south west) rectangle
      ([xshift=1.2cm,yshift=-0cm]current page.north west);
    \node[rotate=90, anchor=center, text width=\paperheight, align=center, font=\bfseries\small, text=gray]
      at ([xshift=0.6cm]current page.west|-current page.center)
      {This version has been peer-reviewed and accepted for publication by Springer}; 
  \end{tikzpicture}%
}

\title{Biomechanics of the abdominal wall before and after ventral hernia repair using dynamic MRI}

\newcommand{\captionMRIpatientsone}{
    Selection of axial and sagittal planes for dynamic acquisitions, example of a pre-operative MRI \\ \textbf{Legend}: \textbf{a)} Axial plane \textbf{b)} Sagittal plane \\ The pink lines represent the respective placements of sagittal and axial planes \\ The white arrow represents the defect dimension measurements (width and height) at rest in axial and sagittal planes respectively \\ The yellow circular arc represents the inter-recti angle (only done on axial MRI), i.e., the angle formed from the aorta/iliac arteries barycenter to the inner tips of rectus abdominis muscles
}
\newcommand{\figureAxialSagittalPlanes}{
    \begin{figure}[H]
        \centering
        \includegraphics[width=0.4\textwidth]{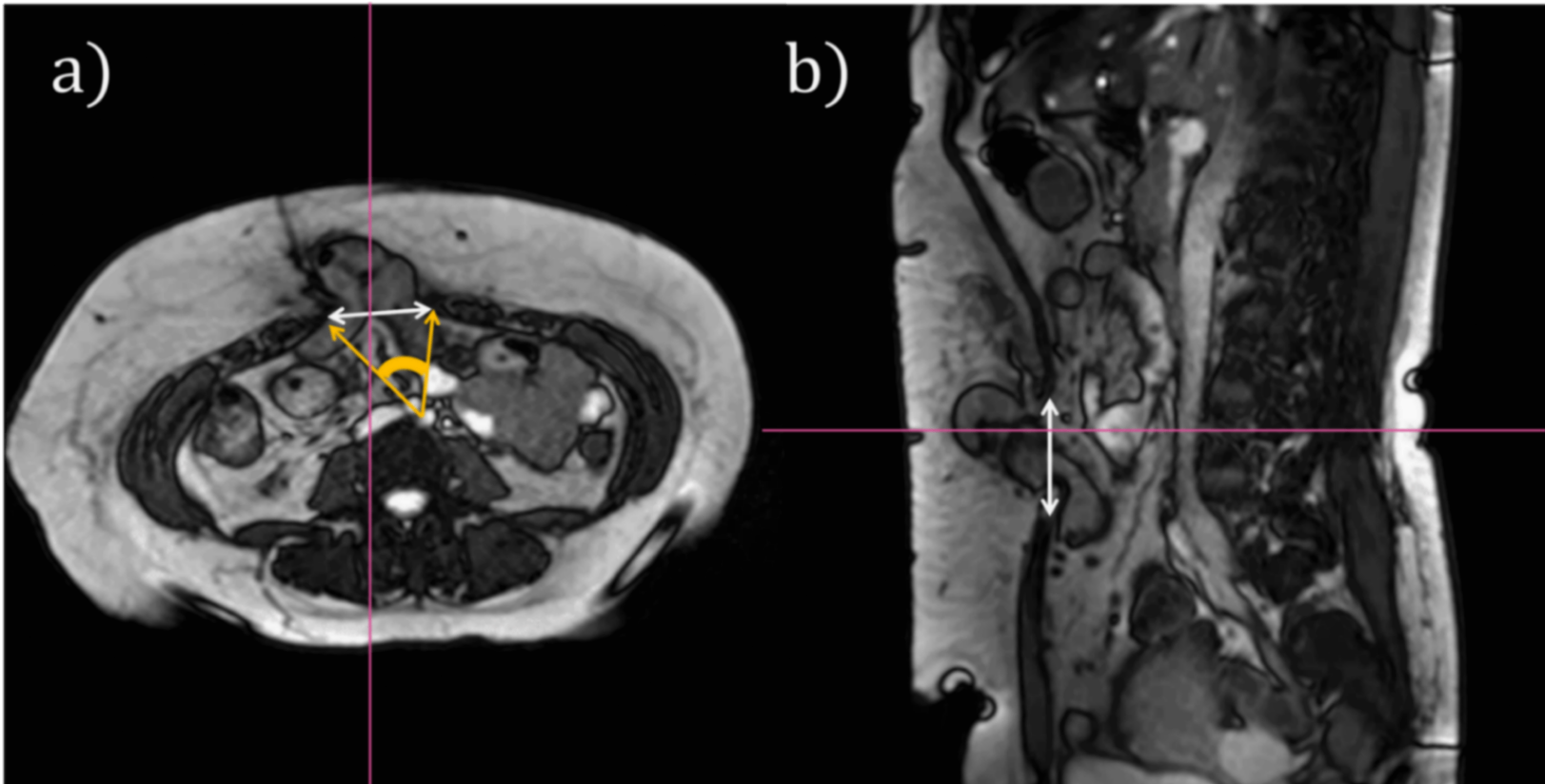}
        \captionsetup{justification=centering, format=plain}
        \caption[Axial and sagittal planes selection for dynamic MRI acquisition]{\captionMRIpatientsone}
        \label{fig:axial_sagittal_planes_onis}
    \end{figure}
}

\newcommand{\captionMRIpatientstwo}{
    Segmentation masks of axial MRI, at rest and during Valsalva contraction \\ \textbf{Legend}: \textbf{a)} Raw MRI \textbf{b)} Hernia sac (for pre-operative stage only) \textbf{c)} Visceral area \textbf{d)} Abdominal muscles: lateral muscles (LM) and rectus abdominis (RA)
}
\newcommand{\figureMasksMRIAxial}{
    \begin{figure}[H]
        \centering
        \includegraphics[width=0.85\textwidth]{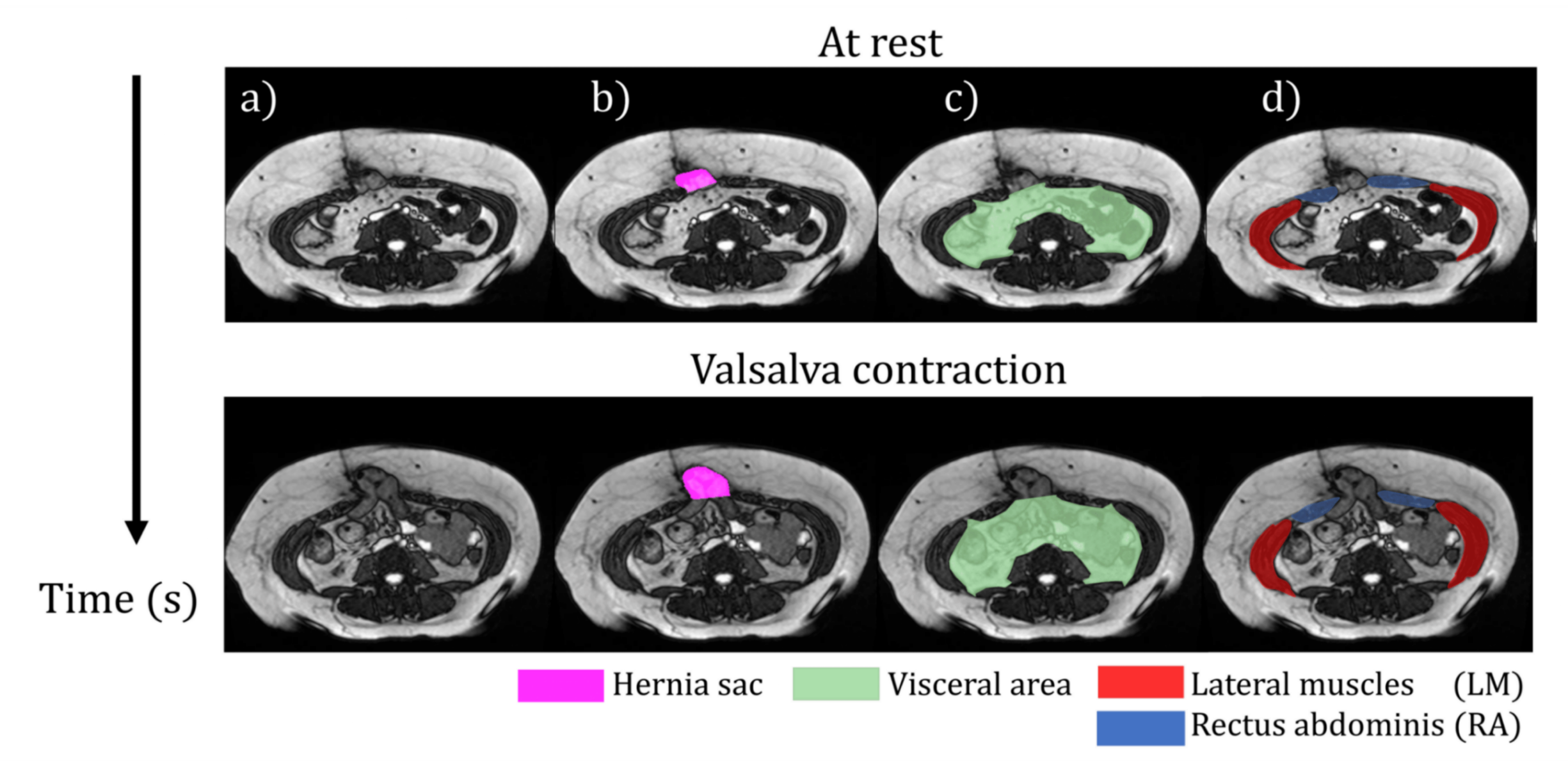}
        \captionsetup{justification=centering, format=plain}
        \caption[Segmentation masks on axial pre-operative MRI]{\captionMRIpatientstwo}
        \label{fig:segmentation_masks_axial}
    \end{figure}
}

\newcommand{\captionMRIpatientsthree}{
    Axial displacement of the abdominal muscles of a preoperative patient performing the Valsalva maneuver \\ The abdominal wall was at rest;  i.e., without contraction; at the beginning and end of the exercise
}
\newcommand{\figureDisplacementTimelineAxial}{
    \begin{figure}[H]
        \centering
        \includegraphics[width=0.99\textwidth]{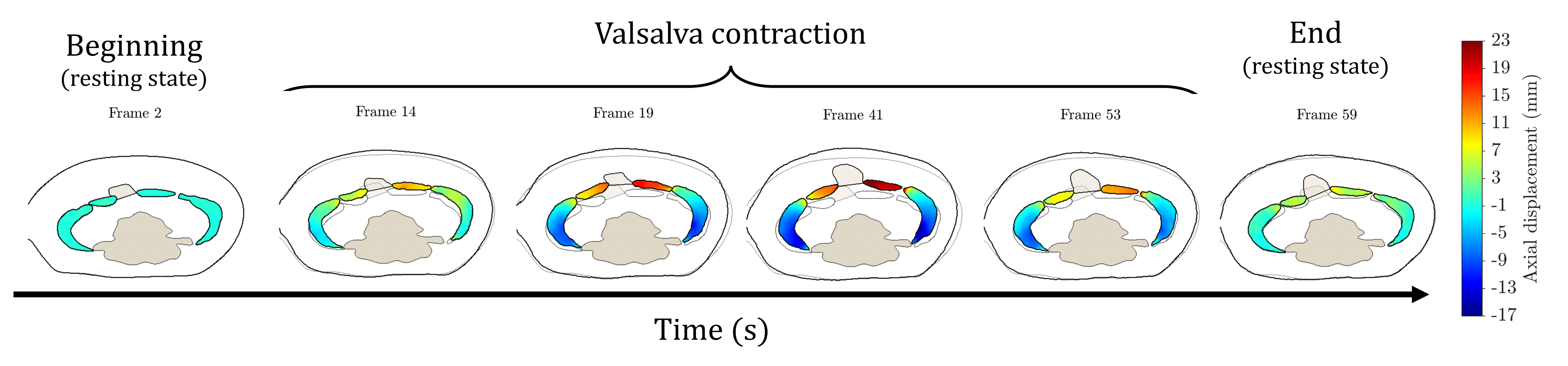}
        \captionsetup{justification=centering, format=plain}
        \caption[Muscle displacement of a pre-operative patient performing a Valsalva maneuver]{\captionMRIpatientsthree}
        \label{fig:displacement_timeline_axial}
    \end{figure}
}

\newcommand{\captionMRIpatientsfour}{
    Segmentation masks in sagittal pre-operative MRI, at rest and during Valsalva contraction \\ \textbf{Legend}: \textbf{a)} Raw MRI \textbf{b)} Hernia sac (for pre-operative stage only) \textbf{c)} Visceral area \textbf{d)} Linea alba segmentation \textbf{e)} Linea alba sections (supra, hernia-scar, infra)
}
\newcommand{\figureMasksMRISagittal}{
    \begin{figure}[H]
        \centering
        \includegraphics[width=0.99\textwidth]{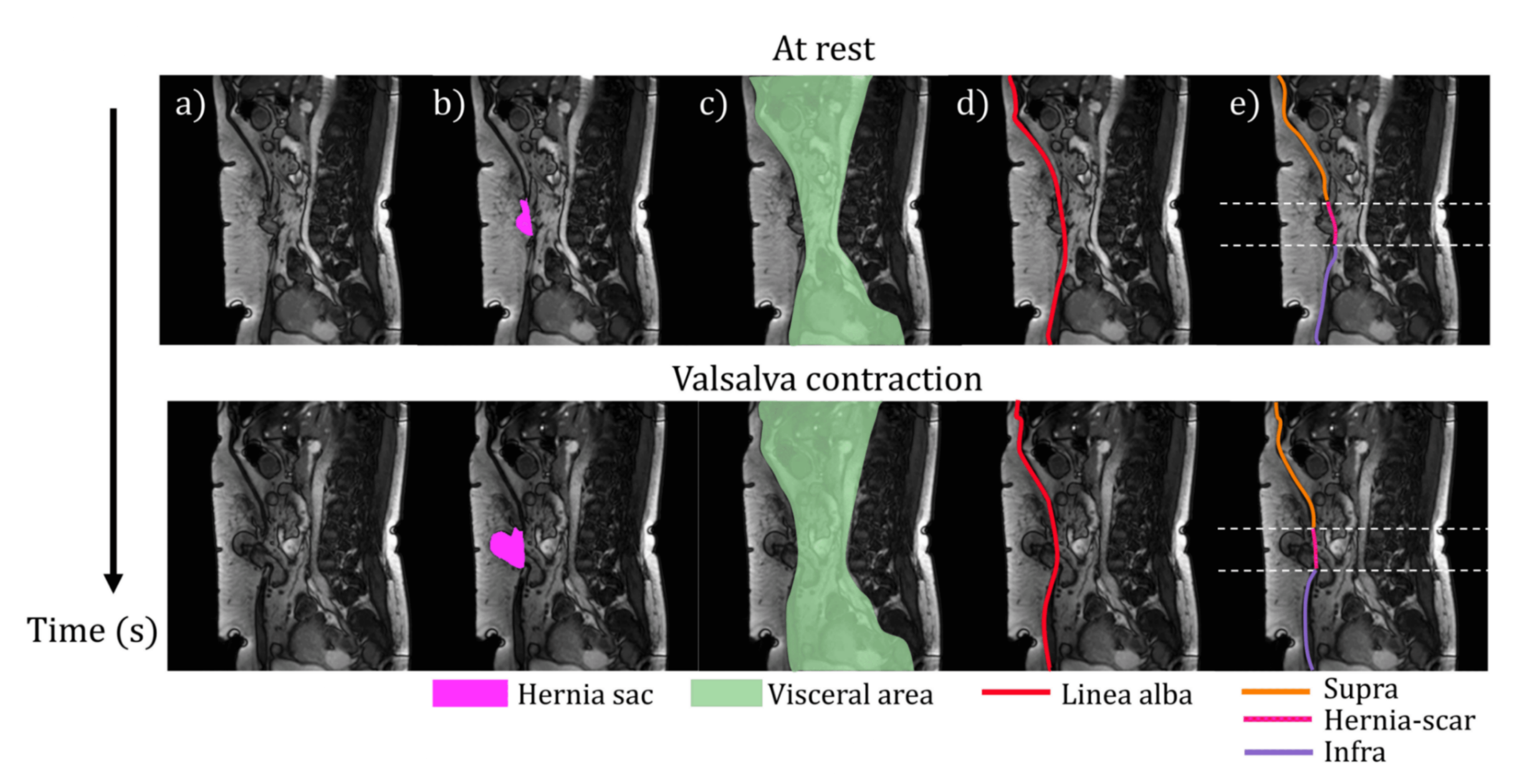}
        \captionsetup{justification=centering, format=plain}
        \caption[Segmentation masks of sagittal pre-operative MRI]{\captionMRIpatientsfour}
        \label{fig:segmentation_masks_sagittal}
    \end{figure}
}

\newcommand{\captionMRIpatientsfive}{
    Sagittal displacement of the linea alba of a preoperative patient performing the Valsalva maneuver \\ The abdominal wall was at rest; i.e., without contraction; at the beginning and end of the exercise
}
\newcommand{\figureDisplacementTimelineSagittal}{
    \begin{figure}[H]
        \centering
        \includegraphics[width=0.99\textwidth]{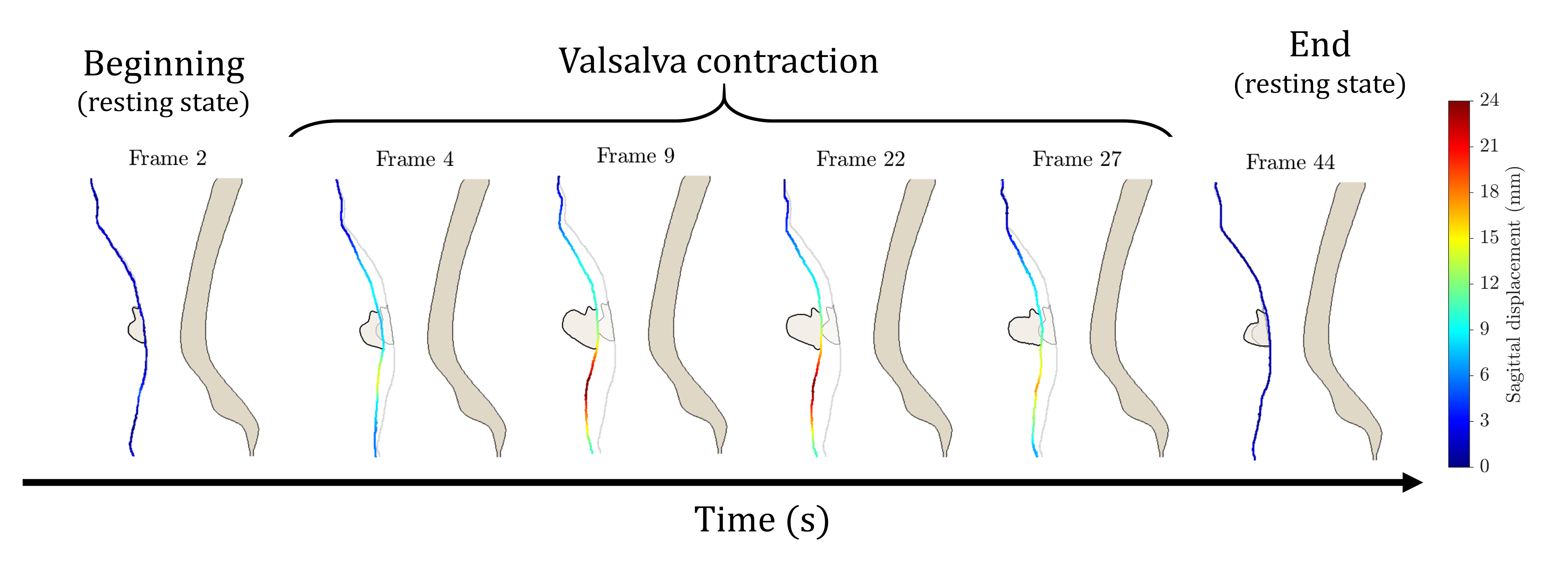}
        \captionsetup{justification=centering, format=plain}
        \caption[Linea alba displacement of a pre-operative patient performing a Valsalva maneuver]{\captionMRIpatientsfive}
        \label{fig:displacement_timeline_sagittal}
    \end{figure}
}

\newcommand{\captionMRIpatientssix}{
    Pre-operative and post-operative inter-recti distance and angle \\ \textbf{Legend}: \textbf{a)} Inter-recti distance averaged among patients (n=10) $\pm$ one standard deviation $\sigma$ \\ \textbf{b)} Corresponding maximum values (i.e., assessed at the end of inhalation for breathing, cough peak for coughing, onset plateau for Valsalva) for each patient \textbf{c)} Inter-recti angle averaged among patients (n=10) \textbf{d)} Corresponding maximum values for each patient
}
\newcommand{\figureInterRectiAngle}{
\begin{figure}[H]
\centering
\includegraphics[width=0.95\textwidth]{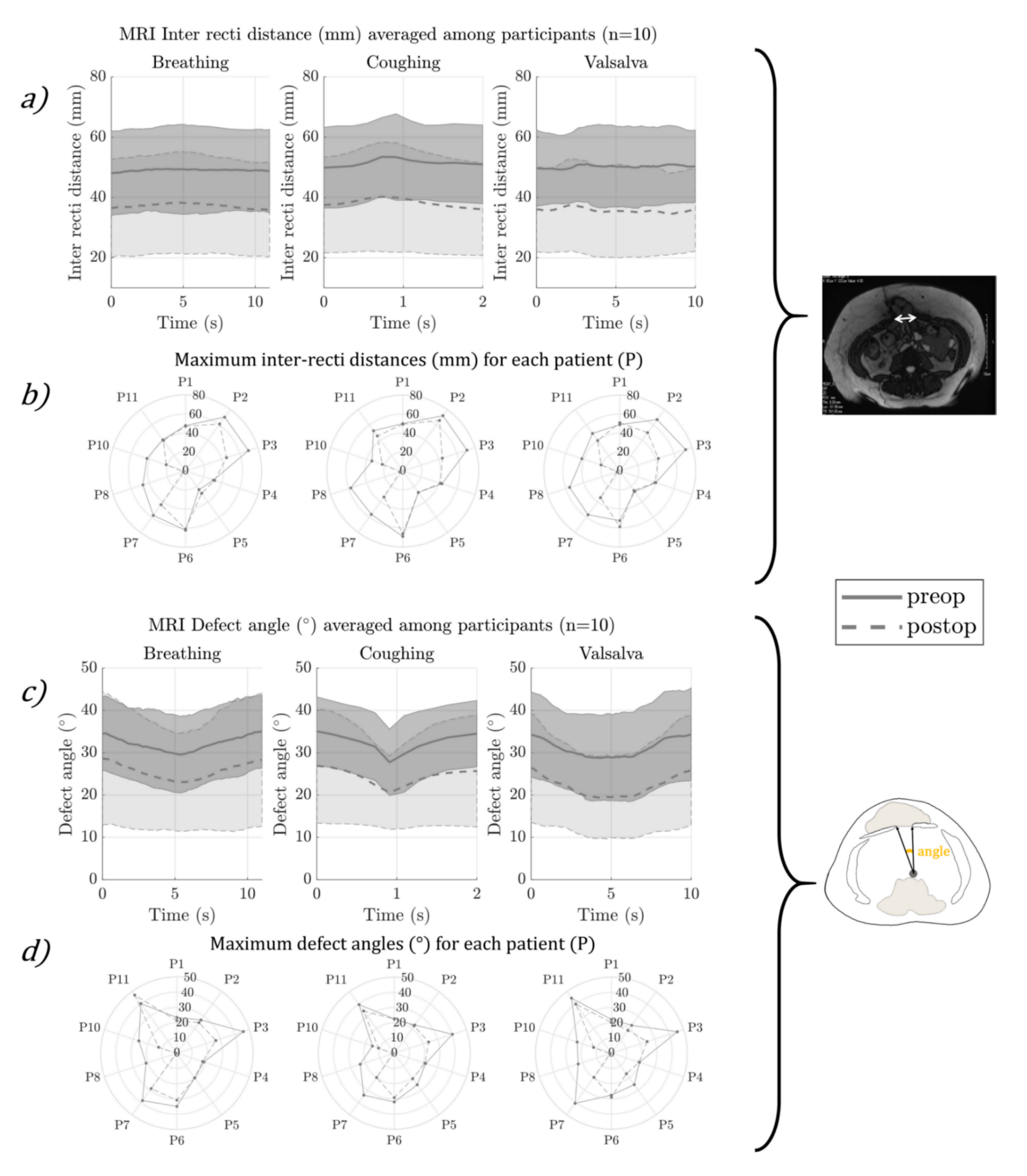}
\captionsetup{justification=centering, format=plain}
\caption[Inter-recti distance and angle before and after surgery]{\captionMRIpatientssix}
\label{fig:Results_diastasis_angle_inter_vol_pre_post}
\end{figure}
}

\newcommand{\captionMRIpatientsseven}{
    Pre-operative and post-operative visceral area \\ \textbf{Legend}: \textbf{a)} Visceral area averaged among patients (n=11) $\pm$ one standard deviation $\sigma$ \\ \textbf{b)} Corresponding maximum values (i.e., assessed at the end of inhalation for breathing, cough peak for coughing, onset plateau for Valsalva) for each patient
}
\newcommand{\figureVisceralArea}{
\begin{figure}[H]
\centering
\includegraphics[width=0.99\textwidth]{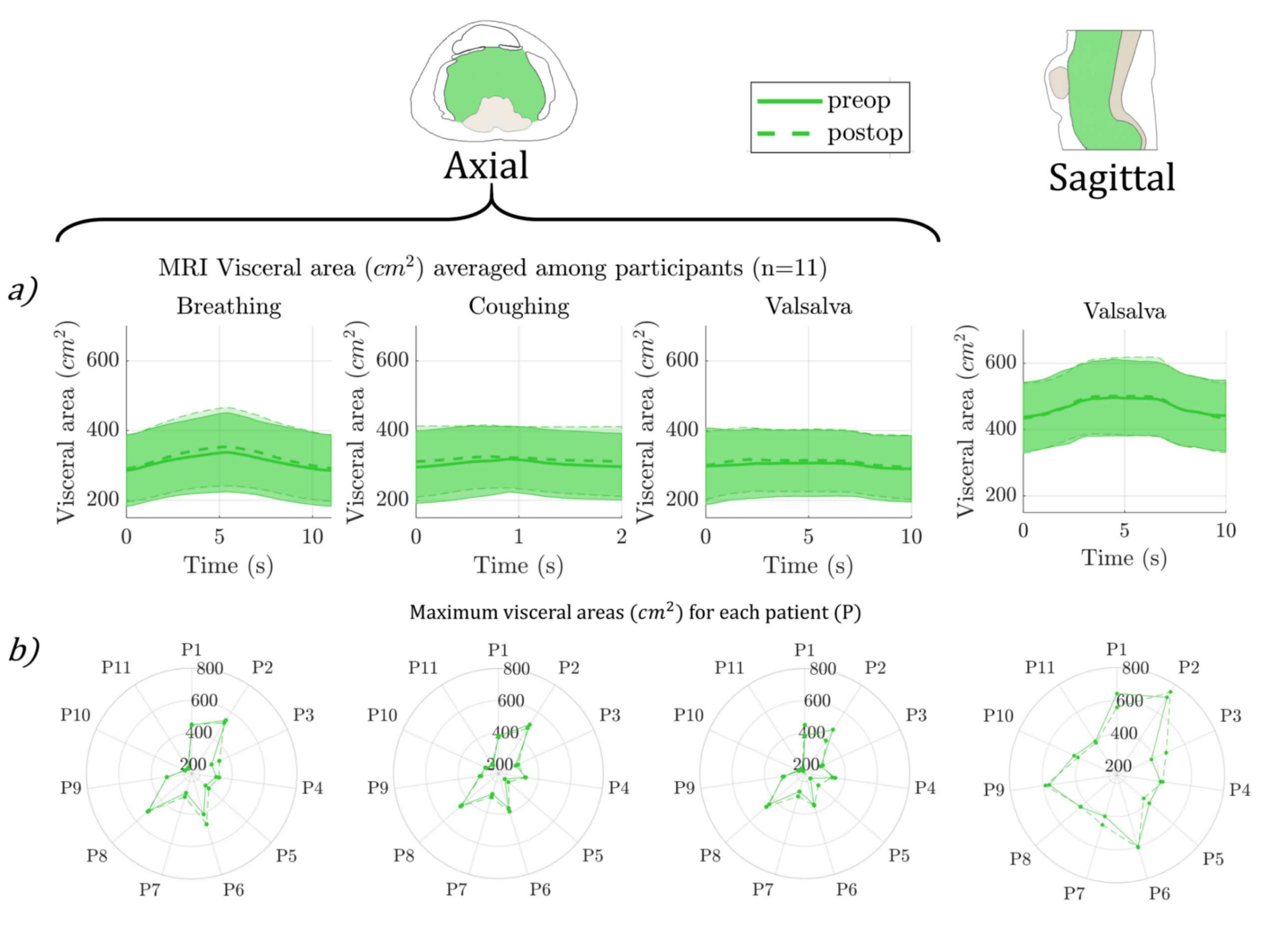}
\captionsetup{justification=centering, format=plain}
\caption[Visceral area before and after hernia surgery]{\captionMRIpatientsseven}
\label{fig:Results_visceral_area_inter_vol_pre_post}
\end{figure}
}

\newcommand{\captionMRIpatientseight}{
    Pre-operative and post-operative axial displacement of lateral muscles (LM) and rectus abdominis (RA) for the three exercises \\ \textbf{Legend}: \textbf{a)} LM displacement averaged among patients (n=11) $\pm$ one standard deviation $\sigma$ \\ \textbf{b)} Corresponding maximum values (i.e., assessed at the end of inhalation for breathing, cough peak for coughing, onset plateau for Valsalva) for each patient \\ \textbf{c)} RA displacement averaged among patients (n=11) $\pm$ one standard deviation $\sigma$ \\ \textbf{d)} Corresponding maximum values for each patient
}
\newcommand{\figureRadialDisp}{
\begin{figure}[H]
\centering
\includegraphics[width=0.99\textwidth]{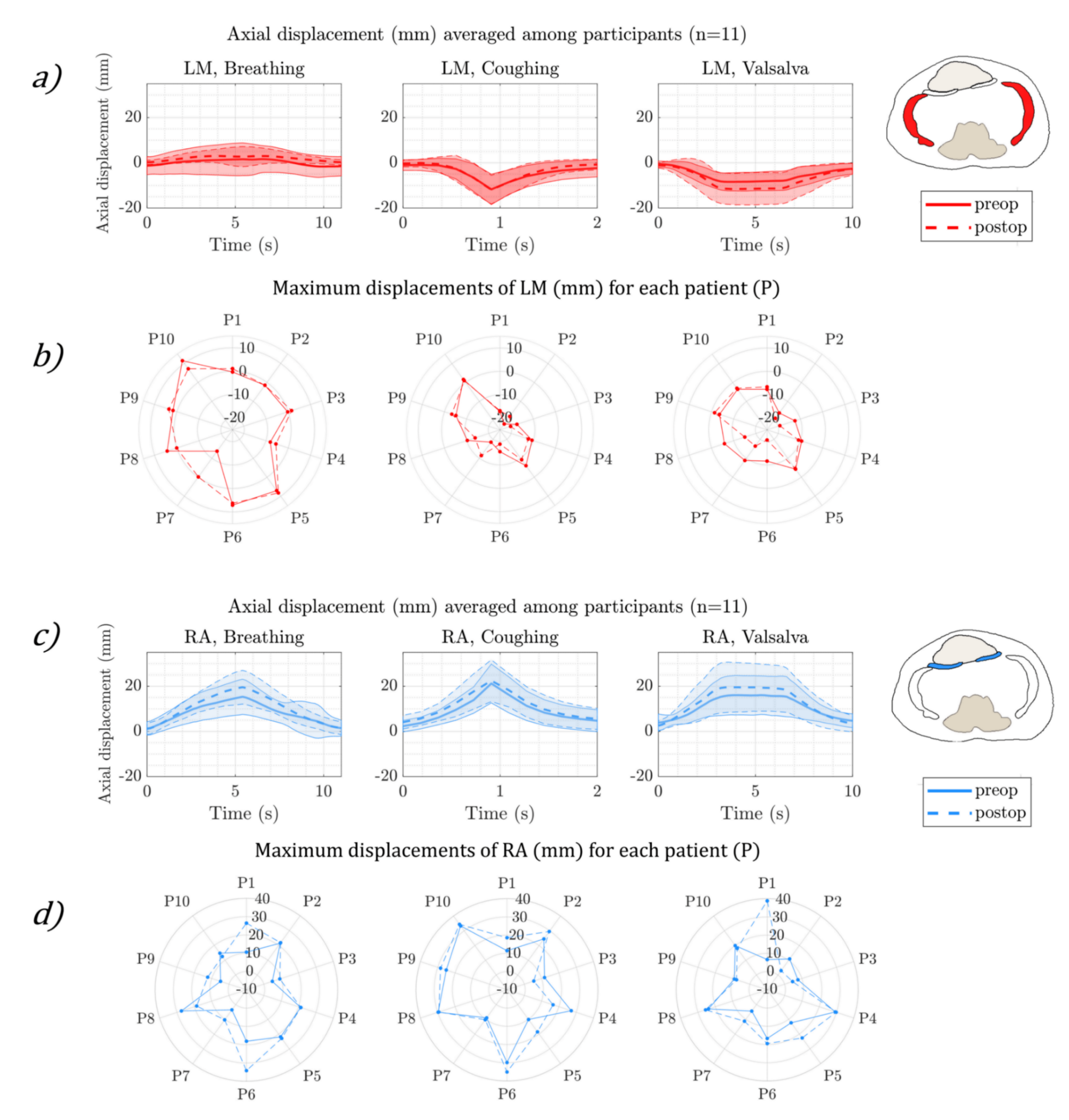}
\captionsetup{justification=centering, format=plain}
\caption[Pre-operative and post-operative muscle displacement on axial MRI]{\captionMRIpatientseight}
\label{fig:Results_radial_disp_inter_vol_pre_post}
\end{figure}
}

\newcommand{\captionMRIpatientsnine}{
    Pre-operative and post-operative sagittal displacement of the 3 regions of linea alba (infra, hernia-scar, supra) during Valsalva \\ \textbf{Legend}: \textbf{a)} Displacement averaged among patients (n=11) \textbf{b)} Corresponding maximum values of each region for each patient \textbf{c)} Corresponding average $\pm$ one standard deviation $\sigma$ for each region
}
\newcommand{\figureLineaAlbaDisp}{
\begin{figure}[H]
\centering
\includegraphics[width=0.99\textwidth]{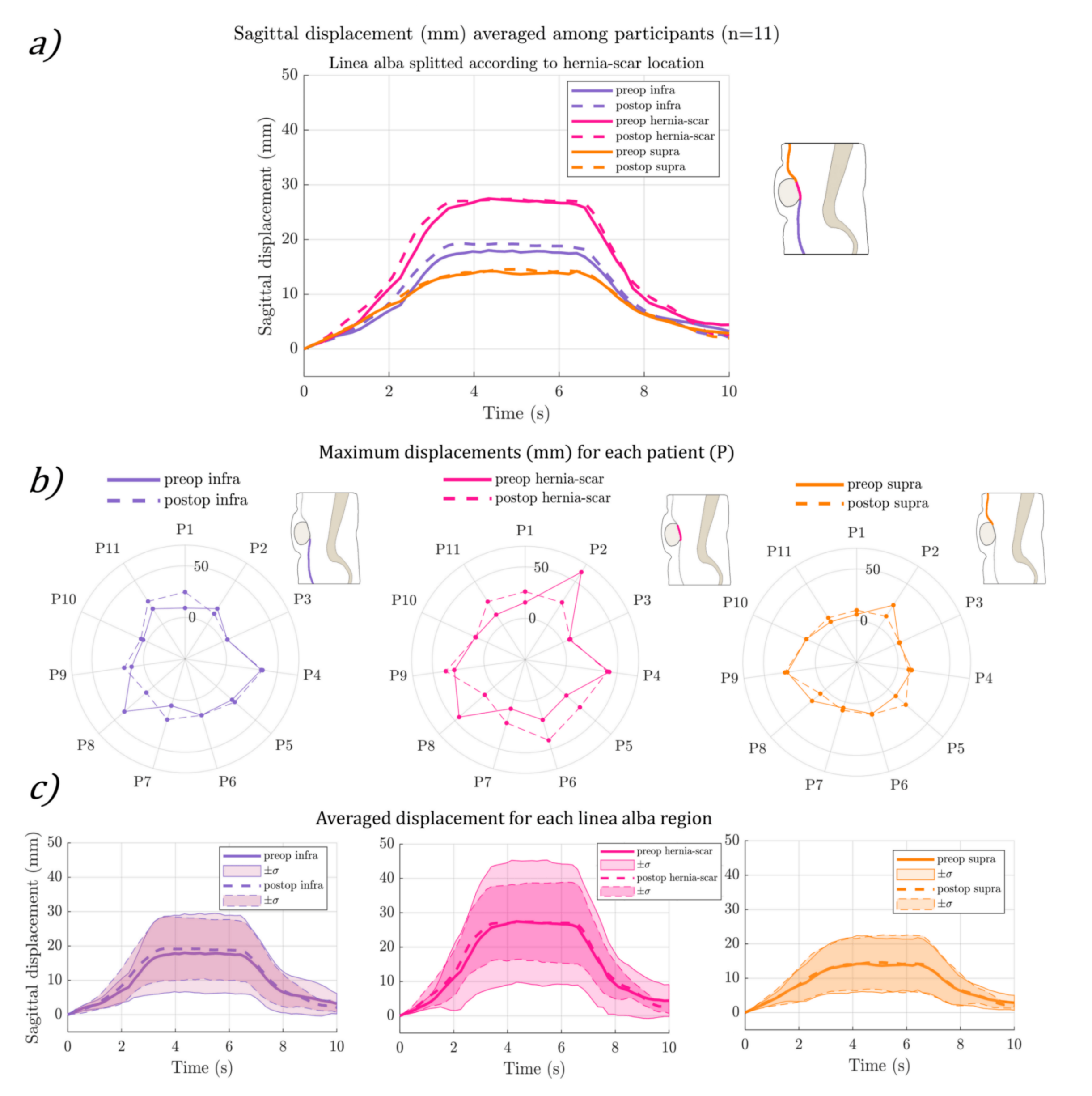}
\captionsetup{justification=centering, format=plain}
\caption[Pre-operative and post-operative linea alba displacement on sagittal MRI]{\captionMRIpatientsnine}
\label{fig:Results_sagittal_disp_inter_vol_pre_post}
\end{figure}
}

\newcommand{\meanBMIPatients}{30.8} 
\newcommand{\stdBMIPatients}{3.1}
\newcommand{\minBMIPatients}{27.7}
\newcommand{\maxBMIPatients}{38.9}

\newcommand{\meanagePatients}{57.9} 
\newcommand{\stdagePatients}{15.1}
\newcommand{\minagePatients}{27.9}
\newcommand{\maxagePatients}{75.2}

\newcommand{\meanHerniaWidth}{42} 
\newcommand{\stdHerniaWidth}{23}
\newcommand{\minHerniaWidth}{15} 
\newcommand{\maxHerniaWidth}{100}

\newcommand{\meanHerniaHeight}{48} 
\newcommand{\stdHerniaHeight}{15}
\newcommand{\minHerniaHeight}{31} 
\newcommand{\maxHerniaHeight}{82}

\newcommand{\meanMeshSurface}{222}
\newcommand{\stdMeshSurface}{161}
\newcommand{\minMeshSurface}{38} 
\newcommand{\maxMeshSurface}{500}

\newcommand{\meandurationpostopmri}{137} 
\newcommand{\stddurationpostopmri}{84}
\newcommand{\mindurationpostopmri}{76}
\newcommand{\maxdurationpostopmri}{362}

\newcommand{\meanFoVaxial}{409} 
\newcommand{\stdFoVaxial}{31}
\newcommand{\minFoVaxial}{360}
\newcommand{\maxFoVaxial}{470}

\newcommand{\meanFoVsagittal}{374} 
\newcommand{\stdFoVsagittal}{28}
\newcommand{\minFoVsagittal}{360}
\newcommand{\maxFoVsagittal}{450}

\newcommand{\meanTRaxial}{160} 
\newcommand{\stdTRaxial}{4}
\newcommand{\minTRaxial}{154}
\newcommand{\maxTRaxial}{166}

\newcommand{\meanTRsagittal}{214} 
\newcommand{\stdTRsagittal}{57}
\newcommand{\minTRsagittal}{156}
\newcommand{\maxTRsagittal}{275}

\newcommand{\meanpXaxial}{0.983} 
\newcommand{\stdpXaxial}{0.075}
\newcommand{\minpXaxial}{0.865}
\newcommand{\maxpXaxial}{1.130}

\newcommand{\meanpXsagittal}{0.898} 
\newcommand{\stdpXsagittal}{0.068}
\newcommand{\minpXsagittal}{0.865}
\newcommand{\maxpXsagittal}{1.082}

\newcommand{\meanRAMpreBreathing}{49.2}
\newcommand{\stdRAMpreBreathing}{15.5}
\newcommand{\meanRAMpostBreathing}{38.2}
\newcommand{\stdRAMpostBreathing}{17.9}

\newcommand{\meanRAMpreCoughing}{53.4}
\newcommand{\stdRAMpreCoughing}{15.1}
\newcommand{\meanRAMpostCoughing}{40}
\newcommand{\stdRAMpostCoughing}{19.2}

\newcommand{\meanRAMpreValsalva}{51.1}
\newcommand{\stdRAMpreValsalva}{13.3}
\newcommand{\meanRAMpostValsalva}{36.5}
\newcommand{\stdRAMpostValsalva}{16.4}

\newcommand{\meanRAMDiffAllExercises}{13}
\newcommand{\stdRAMDiffAllExercises}{16.4}

\newcommand{\meanRatioRAMAllExercises}{26}
\newcommand{\stdRatioRAMAllExercises}{28}

\newcommand{\pvalRAMBreathing}{0.05}
\newcommand{\pvalRAMCoughing}{0.04}
\newcommand{\pvalRAMValsalva}{0.03}

\newcommand{\meanDefectAnglepreBreathing}{29.6}
\newcommand{\stdDefectAnglepreBreathing}{9.6}
\newcommand{\meanDefectAnglepostBreathing}{23}
\newcommand{\stdDefectAnglepostBreathing}{12.2}

\newcommand{\meanDefectAnglepreCoughing}{27.7}
\newcommand{\stdDefectAnglepreCoughing}{8.3}
\newcommand{\meanDefectAnglepostCoughing}{20.6}
\newcommand{\stdDefectAnglepostCoughing}{9.1}

\newcommand{\meanDefectAnglepreValsalva}{29.5}
\newcommand{\stdDefectAnglepreValsalva}{10.1}
\newcommand{\meanDefectAnglepostValsalva}{20.1}
\newcommand{\stdDefectAnglepostValsalva}{10.2}

\newcommand{\meanDefectAngleDiffAllExercises}{7.7}
\newcommand{\stdDefectAngleDiffAllExercises}{8.4}

\newcommand{\pvalDefectAngleBreathing}{0.05}
\newcommand{\pvalDefectAngleCoughing}{0.02}
\newcommand{\pvalDefectAngleValsalva}{0.01}

\newcommand{\meanRadialDispLMpreBreathing}{1.6}
\newcommand{\stdRadialDispLMpreBreathing}{7.6}
\newcommand{\meanRadialDispLMpostBreathing}{2.7}
\newcommand{\stdRadialDispLMpostBreathing}{4.6}

\newcommand{\meanRadialDispLMpreCoughing}{-11.7}
\newcommand{\stdRadialDispLMpreCoughing}{7}
\newcommand{\meanRadialDispLMpostCoughing}{-11.9}
\newcommand{\stdRadialDispLMpostCoughing}{6.7}

\newcommand{\meanRadialDispLMpreValsalva}{-8.2}
\newcommand{\stdRadialDispLMpreValsalva}{4}
\newcommand{\meanRadialDispLMpostValsalva}{-11.2}
\newcommand{\stdRadialDispLMpostValsalva}{7.1}

\newcommand{\meanRadialDispRApreBreathing}{15.3}
\newcommand{\stdRadialDispRApreBreathing}{8.2}
\newcommand{\meanRadialDispRApostBreathing}{19.5}
\newcommand{\stdRadialDispRApostBreathing}{7.8}

\newcommand{\meanRadialDispRApreCoughing}{21.2}
\newcommand{\stdRadialDispRApreCoughing}{9}
\newcommand{\meanRadialDispRApostCoughing}{22.4}
\newcommand{\stdRadialDispRApostCoughing}{9.7}

\newcommand{\meanRadialDispRApreValsalva}{15.3}
\newcommand{\stdRadialDispRApreValsalva}{8.7}
\newcommand{\meanRadialDispRApostValsalva}{19.2}
\newcommand{\stdRadialDispRApostValsalva}{11.6}

\newcommand{\pvalRadialDispRApostMeshSizeBreathing}{0.04}
\newcommand{\pearsonRadialDispRApostMeshSizeBreathing}{-0.65}
\newcommand{\pvalRadialDispRApostMeshSizeCoughing}{0.01}
\newcommand{\pearsonRadialDispRApostMeshSizeCoughing}{-0.78}

\newcommand{\pvalRadialDispLMBreathing}{0.44}
\newcommand{\pvalRadialDispRABreathing}{0.09}
\newcommand{\pvalRadialDispLMCoughing}{0.88}
\newcommand{\pvalRadialDispRACoughing}{0.49}
\newcommand{\pvalRadialDispLMValsalva}{0.047}
\newcommand{\pvalRadialDispRAValsalva}{0.25}

\newcommand{\meanLengthLMpreBreathing}{169.8}
\newcommand{\stdLengthLMpreBreathing}{50.1}
\newcommand{\meanLengthLMpostBreathing}{177}
\newcommand{\stdLengthLMpostBreathing}{50.6}

\newcommand{\meanLengthLMpreCoughing}{143.9}
\newcommand{\stdLengthLMpreCoughing}{48.6}
\newcommand{\meanLengthLMpostCoughing}{151.7}
\newcommand{\stdLengthLMpostCoughing}{44.6}

\newcommand{\meanLengthLMpreValsalva}{146.2}
\newcommand{\stdLengthLMpreValsalva}{52.6}
\newcommand{\meanLengthLMpostValsalva}{155.7}
\newcommand{\stdLengthLMpostValsalva}{53.4}

\newcommand{\meanLengthRApreBreathing}{81.1}
\newcommand{\stdLengthRApreBreathing}{24.4}
\newcommand{\meanLengthRApostBreathing}{90.4}
\newcommand{\stdLengthRApostBreathing}{26.6}

\newcommand{\meanLengthRApreCoughing}{84.7}
\newcommand{\stdLengthRApreCoughing}{23.2}
\newcommand{\meanLengthRApostCoughing}{91.4}
\newcommand{\stdLengthRApostCoughing}{23.9}

\newcommand{\meanLengthRApreValsalva}{82}
\newcommand{\stdLengthRApreValsalva}{23.1}
\newcommand{\meanLengthRApostValsalva}{90.9}
\newcommand{\stdLengthRApostValsalva}{23.1}

\newcommand{\meanThicknessLMpreBreathing}{25.2}
\newcommand{\stdThicknessLMpreBreathing}{3.9}
\newcommand{\meanThicknessLMpostBreathing}{24.6}
\newcommand{\stdThicknessLMpostBreathing}{4.1}

\newcommand{\meanThicknessLMpreCoughing}{31.6}
\newcommand{\stdThicknessLMpreCoughing}{7.3}
\newcommand{\meanThicknessLMpostCoughing}{30.8}
\newcommand{\stdThicknessLMpostCoughing}{7.3}

\newcommand{\meanThicknessLMpreValsalva}{31.4}
\newcommand{\stdThicknessLMpreValsalva}{6}
\newcommand{\meanThicknessLMpostValsalva}{31.6}
\newcommand{\stdThicknessLMpostValsalva}{7}

\newcommand{\meanThicknessRApreBreathing}{13.1}
\newcommand{\stdThicknessRApreBreathing}{1.7}
\newcommand{\meanThicknessRApostBreathing}{13}
\newcommand{\stdThicknessRApostBreathing}{3.3}

\newcommand{\meanThicknessRApreCoughing}{13.5}
\newcommand{\stdThicknessRApreCoughing}{2}
\newcommand{\meanThicknessRApostCoughing}{13}
\newcommand{\stdThicknessRApostCoughing}{3.2}

\newcommand{\meanThicknessRApreValsalva}{14}
\newcommand{\stdThicknessRApreValsalva}{3.5}
\newcommand{\meanThicknessRApostValsalva}{13.5}
\newcommand{\stdThicknessRApostValsalva}{3.9}

\newcommand{\meanRadStrainLMpreBreathing}{-7.9}
\newcommand{\stdRadStrainLMpreBreathing}{13.3}
\newcommand{\meanRadStrainLMpostBreathing}{-11.1}
\newcommand{\stdRadStrainLMpostBreathing}{5.6}

\newcommand{\meanRadStrainLMpreCoughing}{16.1}
\newcommand{\stdRadStrainLMpreCoughing}{19.8}
\newcommand{\meanRadStrainLMpostCoughing}{14.4}
\newcommand{\stdRadStrainLMpostCoughing}{18.4}

\newcommand{\meanRadStrainLMpreValsalva}{11.8}
\newcommand{\stdRadStrainLMpreValsalva}{13.5}
\newcommand{\meanRadStrainLMpostValsalva}{16.7}
\newcommand{\stdRadStrainLMpostValsalva}{18.2}

\newcommand{\meanRadStrainRApreBreathing}{-3.1}
\newcommand{\stdRadStrainRApreBreathing}{8.8}
\newcommand{\meanRadStrainRApostBreathing}{-0.9}
\newcommand{\stdRadStrainRApostBreathing}{8.7}

\newcommand{\meanRadStrainRApreCoughing}{1.3}
\newcommand{\stdRadStrainRApreCoughing}{9.9}
\newcommand{\meanRadStrainRApostCoughing}{-4.7}
\newcommand{\stdRadStrainRApostCoughing}{9.2}

\newcommand{\meanRadStrainRApreValsalva}{4}
\newcommand{\stdRadStrainRApreValsalva}{13.3}
\newcommand{\meanRadStrainRApostValsalva}{0.6}
\newcommand{\stdRadStrainRApostValsalva}{10.9}

\newcommand{\meanCircStrainLMpreBreathing}{6.1}
\newcommand{\stdCircStrainLMpreBreathing}{12.4}
\newcommand{\meanCircStrainLMpostBreathing}{10.4}
\newcommand{\stdCircStrainLMpostBreathing}{9.9}

\newcommand{\meanCircStrainLMpreCoughing}{-12.8}
\newcommand{\stdCircStrainLMpreCoughing}{9.8}
\newcommand{\meanCircStrainLMpostCoughing}{-9.1}
\newcommand{\stdCircStrainLMpostCoughing}{8.3}

\newcommand{\meanCircStrainLMpreValsalva}{-9.2}
\newcommand{\stdCircStrainLMpreValsalva}{7.3}
\newcommand{\meanCircStrainLMpostValsalva}{-6.5}
\newcommand{\stdCircStrainLMpostValsalva}{8.9}

\newcommand{\meanCircStrainRApreBreathing}{2.2}
\newcommand{\stdCircStrainRApreBreathing}{5.9}
\newcommand{\meanCircStrainRApostBreathing}{5.9}
\newcommand{\stdCircStrainRApostBreathing}{4.7}

\newcommand{\meanCircStrainRApreCoughing}{9.6}
\newcommand{\stdCircStrainRApreCoughing}{14.5}
\newcommand{\meanCircStrainRApostCoughing}{6.5}
\newcommand{\stdCircStrainRApostCoughing}{8.5}

\newcommand{\meanCircStrainRApreValsalva}{5.1}
\newcommand{\stdCircStrainRApreValsalva}{8.3}
\newcommand{\meanCircStrainRApostValsalva}{5.9}
\newcommand{\stdCircStrainRApostValsalva}{8.2}

\newcommand{\meanRatioSagDispInfraValsalva}{60.2}
\newcommand{\stdRatioSagDispInfraValsalva}{67}

\newcommand{\meanRatioSagDispHerniaValsalva}{69.3}
\newcommand{\stdRatioSagDispHerniaValsalva}{53.6}

\newcommand{\meanRatioSagDispSupraValsalva}{42.3}
\newcommand{\stdRatioSagDispSupraValsalva}{38.7}

\newcommand{\pvalRatioSagDispInfraHerniaValsalva}{0.88}
\newcommand{\pvalRatioSagDispInfraSupraValsalva}{0.31}
\newcommand{\pvalRatioSagDispHerniaSupraValsalva}{0.07}

\newcommand{\meanHerniaSacpreBreathing}{21.5}
\newcommand{\stdHerniaSacpreBreathing}{26.4}
\newcommand{\meanHerniaSacpreCoughing}{21.8}
\newcommand{\stdHerniaSacpreCoughing}{23.3}
\newcommand{\meanHerniaSacpreValsalva}{22.6}
\newcommand{\stdHerniaSacpreValsalva}{21.1}

\newcommand{\meanHerniaSacVariationpreBreathing}{17.3}
\newcommand{\stdHerniaSacVariationpreBreathing}{57.4}
\newcommand{\meanHerniaSacVariationpreCoughing}{128.4}
\newcommand{\stdHerniaSacVariationpreCoughing}{199.2}
\newcommand{\meanHerniaSacVariationpreValsalva}{35}
\newcommand{\stdHerniaSacVariationpreValsalva}{44}

\newcommand{\meanSagHerniaSacVariationpreValsalva}{33.6}
\newcommand{\stdSagHerniaSacVariationpreValsalva}{50.8}

\newcommand{\meanSagHerniaSacpreValsalva}{19.1}
\newcommand{\stdSagHerniaSacpreValsalva}{11.4}

\newcommand{\meanVisceralAreapreBreathing}{337.9}
\newcommand{\stdVisceralAreapreBreathing}{118.9}
\newcommand{\meanVisceralAreapreCoughing}{318.5}
\newcommand{\stdVisceralAreapreCoughing}{99.3}
\newcommand{\meanVisceralAreapreValsalva}{305.6}
\newcommand{\stdVisceralAreapreValsalva}{101.3}

\newcommand{\meanVisceralAreapostBreathing}{353.9}
\newcommand{\stdVisceralAreapostBreathing}{118.3}
\newcommand{\meanVisceralAreapostCoughing}{323.4}
\newcommand{\stdVisceralAreapostCoughing}{92.9}
\newcommand{\meanVisceralAreapostValsalva}{315.2}
\newcommand{\stdVisceralAreapostValsalva}{93.9}

\newcommand{\pvalVisceralAreaBreathing}{0.06}
\newcommand{\pvalVisceralAreaCoughing}{0.36}
\newcommand{\pvalVisceralAreaValsalva}{0.45}

\newcommand{\meanSagVisceralAreapreValsalva}{490.6}
\newcommand{\stdSagVisceralAreapreValsalva}{115.9}
\newcommand{\meanSagVisceralAreapostValsalva}{496.5}
\newcommand{\stdSagVisceralAreapostValsalva}{115.9}

\newcommand{\pvalSagVisceralAreaValsalva}{0.7}

\newcommand{\PvaluethresholdPatients}{0.05}

\author[1,2]{Victoria Joppin \orcidlink{0000-0003-2789-6647}}
\author[2]{David Bendahan \orcidlink{0000-0002-1502-0958}}
\author[1,3]{Ahmed Ali El Ahmadi \orcidlink{0000-0002-4209-2113}}
\author[1]{Catherine Masson \orcidlink{0000-0003-3578-9067}}
\author[1,4]{Thierry Bege \orcidlink{0000-0002-0775-3035}}

\affil[1]{Aix Marseille Univ, Univ Gustave Eiffel, LBA, Marseille France}
\affil[2]{Aix Marseille Université, CNRS, CRMBM UMR 7339, Marseille France}
\affil[3]{Department of Radiology, North Hospital, Assistance Publique Hôpitaux de Marseille, Marseille, France}
\affil[4]{Department of General Surgery, Aix-Marseille Univ, North Hospital, APHM, Marseille, France}

\begin{document}

\date{Publication date: May 24th, 2025}

\maketitle

\section*{Authors}

\noindent Victoria Joppin \orcidlink{0000-0003-2789-6647} \tab \tab victoria.joppin@proton.me

\noindent David Bendahan \orcidlink{0000-0002-1502-0958} \tab david.bendahan@univ-amu.fr

\noindent Ahmed Ali El Ahmadi \orcidlink{0000-0002-4209-2113} \tab ahmed.el-ahmadi@ap-hm.fr

\noindent Catherine Masson \orcidlink{0000-0003-3578-9067} \tab catherine.masson@univ-eiffel.fr

\noindent Thierry Bege \orcidlink{0000-0002-0775-3035} \tab \tab thierry.bege@ap-hm.fr

\subsection*{Corresponding author}

\noindent Victoria Joppin \orcidlink{0000-0003-2789-6647} \tab victoria.joppin@proton.me

\noindent Laboratoire de Biomécanique Appliquée, UMRT24 Université Gustave Eiffel - Aix Marseille Université

\noindent Faculté des Sciences Médicales et Paramédicales - Secteur Nord

\noindent 51 Boulevard Pierre Dramard, F-13016 Marseille, France

\newpage

\section*{Abstract}

\noindent \textit{Purpose}: This study aims to investigate the use of dynamic MRI to assess abdominal wall biomechanics before and after hernia surgery, considering that such evaluations can enhance our understanding of physiopathology and contribute to reducing recurrence rates.

\vspace{0.2cm}

\noindent \textit{Methods}: Patients were assessed using dynamic MRI in axial and sagittal planes while performing exercises (breathing, coughing, Valsalva) before and after their abdominal hernia surgery with mesh placement. Rectus and lateral muscles, linea alba, viscera area, defect dimensions and hernia sac were contoured with semi-automatic process to quantify the abdominal wall biomechanical temporal modifications.

\vspace{0.2cm}

\noindent \textit{Results}: This study enrolled 11 patients. During coughing, the axial area of the hernia sac increased by $\meanHerniaSacVariationpreCoughing$ $\pm$ $\stdHerniaSacVariationpreCoughing$$\%$. The sac increased similarly in axial and sagittal planes during Valsalva. Post-surgical evaluations showed a $\meanRatioRAMAllExercises\%$ reduction in inter-recti distance and a lengthening of all muscles (\textit{p} $\leq$ $\PvaluethresholdPatients$). The post-operative rectus abdominis thickness change was negatively correlated with defect width during breathing (\textit{p} $\leq$ $\PvaluethresholdPatients$). The largest change in linea alba displacement was observed in the surgical site (\textit{p} = $\pvalRatioSagDispHerniaSupraValsalva$). Post-operatively, lateral muscles had a larger inward displacement during Valsalva (\textit{p} $\leq$ $\PvaluethresholdPatients$). Rectus abdominis had a larger outward displacement during breathing (\textit{p} = $\pvalRadialDispRABreathing$), reduced with the mesh size (\textit{p} $\leq$ $\PvaluethresholdPatients$). A large inter-individual variability was observed.

\vspace{0.2cm}

\noindent \textit{Conclusion}: Using a semi-automatic methodology, an in-depth analysis of the biomechanics of the abdominal wall was conducted, highlighting the importance of a patient-specific assessment. A broader study and consideration of recurrence would subsequently complete this methodological work.

\vspace{1cm}

\noindent \textbf{Keywords}

\noindent Hernia surgery; Abdominal wall function; Dynamic MRI; Biomechanics; Surgical outcomes; Medical imaging post-processing

\vspace{0.5cm}

\noindent \textbf{Abbreviations}

\noindent LM: lateral muscles; RA: rectus abdominis; MRI: magnetic resonance imaging; IPOM: Intra-Peritoneal Onlay Mesh; CT: computed tomography; \textit{p}: \textit{p}-value; \textit{r}: Pearson correlation coefficient

\newpage

\section{Introduction}

\noindent Abdominal wall hernias are characterised by the protrusion of abdominal contents through a weakened abdominal wall. They are common pathologies and associated with significant health and economic burdens \cite{schlosserVentralHerniaRepair2023}.

\vspace{0.5cm}

\noindent Hernia surgery success is traditionally measured by recurrence and complication rates. Although improvements in surgical techniques and meshes \cite{lopez-canoIncisionalHerniaDepends2024,bhardwajYearOverYearVentralHernia2024,meijerPrinciplesAbdominalWound2016} has been shown to improves the success of ventral incisional hernia repair \cite{mussackHealthrelatedQualityoflifeChanges2006,smithHealthrelatedQualityLife2022}, the recurrence rate is still very high reaching up to 45$\%$ after 5-year post-surgery \cite{bhardwajYearOverYearVentralHernia2024}. These disappointing results must be addressed through new approaches aimed at objectively and quantitatively evaluating abdominal wall function before and after hernia surgery \cite{jensenAbdominalMuscleFunction2014}. 

\vspace{0.5cm}

\noindent Previous \textit{in vivo} studies have assessed the impact of hernia on muscle functionality. The corresponding investigative methods were surface electromyography and dynamometry. An electromyography-based study revealed that abdominal muscle strength was significantly reduced in patients with incisional hernia compared to healthy \cite{royComparativeStudyEvaluate2023}. Various studies using isokinetic dynamometer \cite{strigardGiantVentralHernia2016,garciamorianaEvaluationRectusAbdominis2023} and strain gauge \cite{sanchezarteagaImpactIncisionalHernia2024} showed a negative correlation between abdominal wall muscle strength and hernia defect width measured on computed tomography (CT) scans. Abdominal muscle strength was also negatively correlated with abdominal rectus diastasis width measured under the umbilicus on CT scan before their diastasis surgery \cite{gunnarssonCorrelationAbdominalRectus2015}.
\noindent Abdominal wall functionality following hernia repair with linea alba restoration was shown to be improved through isokinetic and isometric measurements of the rectus muscle \cite{crissFunctionalAbdominalWall2014}. Preoperative core training was found to significantly enhance muscle strength after ventral hernia repair \cite{ahmedEffectPreoperativeAbdominal2018}.

\noindent While these studies provide valuable insights, they are often limited by their reliance on static, strength-based measurements. Assessment of abdominal wall anatomy and biomechanics during physiological activities are still very scarce and dynamic imaging methods might be of high interest in this regard.

\vspace{0.5cm}

\noindent Ultrasound and CT scan imaging techniques are frequently used for anatomical investigation before hernia surgery. Magnetic resonance imaging (MRI) is less used in clinical evaluation but may have some specific interest \cite{plumbContemporaryImagingRectus2021}. Ultrasound can be used to diagnose hernias \cite{qiuMeasurementsAbdominalWall2023} and help in understanding the etiology of incisional hernia formation \cite{harlaarDevelopmentIncisionalHerniation2017}. Ultrasound has been shown to be useful in the dynamic assessment of abdominal adhesions \cite{yaseminAssessmentDiagnosticEfficacy2020}. However, it only provides a local analysis, and the abdominal image quality has been shown to be negatively affected by patients' body mass indexes (BMI) \cite{jeejiIncreasedBodyMass2021}. CT scan is used for hernia detection \cite{muysomsEuropeanHerniaSociety2015}, anatomical evaluation for surgery planning, and post-operative follow-up \cite{gutierrezdelapenaValueCTDiagnosis2001,gossiosValueCTLaparoscopic2003}. Changes in abdominal wall morphometry induced by hernia surgery have been mainly investigated using CT scan \cite{lisieckiAbdominalWallDynamics2015}. A decreased cross-sectional anterior-posterior distance, area and circumference of the abdomen have been reported after surgery.
It has also been suggested that a more functional-based preoperative CT scan imaging of the abdominal wall during Valsalva maneuver could markedly enhance the comprehension of ventral hernia defects and facilitate the design of efficient surgical plans \cite{readImagingInsightsAbdominal2022,kallinowskiAssessingGRIPVentral2018,kallinowskiCTAbdomenValsalva2019}.

\noindent However, different studies showed that CT scan was not associated with reliable diagnosis in ventral hernia recurrence \cite{holihanUseComputedTomography2016}. In addition, CT reports often fail to include important pre-operative hernia features \cite{kushnerIdentifyingCriticalComputed2021}. One has to keep in mind that CT scan is a radiating tool offering static views that fail to capture the dynamic nature of hernias during physiological activities such as breathing and coughing. In that respect, MRI appears to be a tool of interest regarding its non-radiating nature and its potential for an in-depth and dynamic study of the whole abdominal wall. Dynamic magnetic resonance imaging (MRI), also called cine-MRI, represents a significant advancement in this field. It provides real-time insights into the movement and interaction of abdominal structures \cite{randallNovelDiagnosticAid2017}. It has multiple applications, such as the quantification of cross-sectional abdominal wall motion \cite{jourdanDynamicMRIQuantificationAbdominal2022} and its relationship with intra-abdominal pressure \cite{joppinBetterUnderstandingAbdominal2025}, abdominal adhesions assessment \cite{yaseminAssessmentDiagnosticEfficacy2020}, study of the effect of wearing an abdominal binder during contraction \cite{miyamotoFastMRIUsed2002}, study of chronic groin pain in inguinal hernias \cite{dallaudiereDynamicMagneticResonance2024} or visualisation of the implanted mesh \cite{fischerFunctionalCineMRI2007,ciritsisTimeDependentChangesMagnetic2014}.

\vspace{0.5cm}

\noindent The present study aims to fill the gap in the objective and functional evaluation of the abdominal wall by investigating anatomical and biomechanical changes in hernia patients before and after surgery. Using dynamic MRI during various exercises, this study intends to provide a real-time, \textit{in vivo} assessment of abdominal wall function, offering a more detailed understanding of post-surgical outcomes.
\section{Methods}
\subsection{Participants}

\noindent This study was approved by the French ethics committee (IDRCB: 2021-A02119-32) according to national legislation related to interventional research and the Declaration of Helsinki.

\vspace{0.2cm}

\noindent Participants accepted to take part in the study after providing their informed written consent. They had to meet the following inclusion criteria: being of legal age, having abdominal wall pathology (primary or incisional hernias) requiring laparotomy surgery at the time of the study, having a defect of 10 cm maximum, having a social security coverage. Exclusion criteria include pregnant or breast-feeding women, people under guardianship, people requiring emergency hernia/ventricular surgery, people with an active digestive pathology other than hernia, the usual contraindications for an MRI examination (claustrophobia, magnetized foreign bodies).

\vspace{0.2cm}

\noindent The hernias were classified according to the EHS classification \cite{muysomsClassificationPrimaryIncisional2009}. The hernia surgery technique was determined by the surgeon, performed by laparotomy or laparoscopy, including a systematic defect closure with mesh placement. Correction of the diastasis was not systematically performed.

\subsection{MRI protocol}

\noindent Participants had a pre-operative MRI and get their hernia surgery. Once their state of healing and pain allowed it, participants underwent post-operative MRI. Participants were positioned in a supine position within the 3-Tesla MRI scanner (MAGNETOM Vida, Siemens Healthineers, Erlangen, Germany). The radiofrequency coils used were those integrated within the bed scanner and a flexible body coil placed on the abdomen. This coil was not attached to the bed so as to ensure a non-restricted anteroposterior movement of the abdomen.

\vspace{0.2cm}

\noindent First, two volumes using 2D sequences T2 HASTE (5mm jointed slices) and 3D sequence T1 DIXON (3mm) were recorded during apnea after inhalation (from the pubis to the xiphoid process) in axial plane. These volumes were then used to position the axial and sagittal planes for the subsequent dynamic acquisitions. Dynamic sequences used were True FISP sequences (balanced gradients sequences) with one slice of 8 mm repeated during 40 to 60 sec. For the pre-operative MRI, dynamic axial acquisitions were performed over the cross-sectional plane where the hernia neck was the largest, as identified from the 3D volumes. The sagittal plane was located at the midpoint of the hernia neck, as illustrated in \autoref{fig:axial_sagittal_planes_onis}. For the post-operative MRI, the axial and sagittal planes were selected to be as similar as possible to those chosen for the preoperative MRI.

\noindent Participants performed three audio-guided exercises after a pre-training session, and each exercise was repeated four times to improve reproducibility. These activities could be done in daily life: deep breathing for 10 seconds, coughing, and performing the Valsalva maneuver, which involves forcing air out against closed mouth and nostrils for 8 seconds. 

\vspace{0.2cm}

\noindent The MRI acquisition parameters, including field of view (window size) and temporal resolution (timestep between two images), were tailored to the individual participant based on their body size. The MRI acquisition parameters can be found in \autoref{appendix:appendix_MRI_parameters}.

\figureAxialSagittalPlanes

\subsection{Data processing}

\noindent MRI scans were initially converted from DICOM to Nifti format using the \textit{dcm2niix} software \cite{liFirstStepNeuroimaging2016}. MR images were visualised using the ImageJ software. MRI sequence of each patient and each exercise was truncated to include only images during exercise execution, beginning and ending at abdominal wall's resting state. The first image was selected by an operator as the slice without visible contraction, representing the initial resting state. The last slice was identified using both subjective visual assessment of the same operator and objective similarity measurements using MedPy, a Python medical image processing library \cite{maierMedPyMedicalImage}. Objective assessments included the Dice Similarity Coefficient (the commonly used metric quantifying spatial overlap between two images \cite{zouStatisticalValidationImage2004}) and mutual information (a statistical measure of shared information between images \cite{maesMedicalImageRegistration2003}). The last slice was selected as the one most closely resembling the first, ensuring the sequence ended with the participant returning to the resting state.

\subsubsection{Axial MRI}

\noindent Each patient had dynamic axial acquisition during which they performed audio-guided breathing, coughing and Valsalva maneuver. 

\noindent For the pre-operative MRI, the defect width was measured at the resting state (\autoref{fig:axial_sagittal_planes_onis}). The hernia sac was delineated whenever feasible (\autoref{fig:segmentation_masks_axial}-b) and the corresponding area ($cm^2$) was quantified. In relative terms, the area variation with respect to the resting state ($area_0$) was also computed, i.e., $area-area_0/area_0$ expressed in percentage ($\%$).

\noindent As illustrated in \autoref{fig:segmentation_masks_axial}-c, the visceral area was defined as the region enclosed by the internal surface of the abdominal muscles and the dorsal area, which includes the quadratus lumborum, psoas major, erector spinae muscles, vertebral discs and body, as previously described \cite{joppinBetterUnderstandingAbdominal2025}.

\noindent Additionally, the rectus abdominis (RA) and lateral muscles (LM), which include the transversus abdominis, internal obliques, and external obliques, were delineated in the MR images using FSLeyes, the image viewer included in the FSL toolbox (\autoref{fig:segmentation_masks_axial}-d. The segmentation method previously described by Ogier \textit{et al.} \cite{ogierIndividualMuscleSegmentation2017} was used to semi-automatically segment the corresponding muscles within the region of interest.

\figureMasksMRIAxial

\noindent In both pre and post-operative images, the distance between the two inner tips of the rectus abdominis muscles was measured and defined as the inter-recti distance. Additionally, the abdominal aorta or iliac arteries were segmented in order to compute its barycenter. Then, the inter-recti angle (in degree) was computed. It was defined as the angle between the inter-recti distance and the aorta/iliac arteries barycenter, as represented in yellow in \autoref{fig:axial_sagittal_planes_onis}. The inter-recti angle reflects both the anatomical separation of the rectus muscles (how wide apart they are) and their positional relationship to the central axis of the body, represented by the aorta/iliac arteries.

\vspace{0.5cm}

\noindent As illustrated in \autoref{fig:displacement_timeline_axial} for the Valsalva exercise, the axial displacement of abdominal muscles was calculated according to the methodology proposed by Jourdan \textit{et al.} \cite{jourdanDynamicMRIQuantificationAbdominal2022}. This displacement is expressed in the radial direction and was subject to change at each time point.

\vspace{0.5cm}

\noindent Circumferential and radial strains respectively noted $\varepsilon_{\theta}$ and $\varepsilon_r$ were computed for both LM and RA muscles \cite{jourdanDynamicMRIQuantificationAbdominal2022}. These variables were defined respectively for the considered group (LM or RA) as the medial axis length and thickness variation with respect to the resting state, i.e., $length-length_0/length_0$ for $\varepsilon_{\theta}$, and $thickness-thickness_0/thickness_0$ for $\varepsilon_r$, expressed in percentage ($\%$).

\figureDisplacementTimelineAxial

\noindent Given that each exercise was performed multiple times by each participant, the different repetitions were then averaged using Matlab R2024a software, in order to obtain a representative average cycle of each exercise for each patient for the two abdominal wall conditions before and after the surgery, as previously described \cite{joppinBetterUnderstandingAbdominal2025}.

\subsubsection{Sagittal MRI}

\noindent For the sagittal dynamic acquisition, the post-processing has been focused on Valsalva exercise. The field of view (lower and upper borders) for pre-operative and post-operative MRI scans was aligned vertically using the position of the L4 vertebra, to ensure the same field of view for both stages.

\noindent As with the axial pre-operative MRI, the defect height was measured at the resting state (\autoref{fig:axial_sagittal_planes_onis}) and hernia sac was segmented whenever feasible, as illustrated \autoref{fig:segmentation_masks_sagittal}-b.

\noindent The visceral area, shown in \autoref{fig:segmentation_masks_sagittal}-c was defined as the region enclosed by the internal surface of the linea alba and the vertebral arch.

\noindent The linea alba, illustrated in \autoref{fig:segmentation_masks_sagittal}-d, was delineated in the sagittal MR images using the same semi-automatic segmentation method employed for axial MRI \cite{ogierIndividualMuscleSegmentation2017}.

\figureMasksMRISagittal

\noindent As shown in \autoref{fig:displacement_timeline_sagittal}, the sagittal displacement within the segmented mask of the linea alba was computed. This displacement was computed in a cartesian coordinate system and evolved at each timestep. The linea alba mask was divided into discrete regions for each patient in order to facilitate assessment within specific areas.

\noindent As illustrated in \autoref{fig:segmentation_masks_sagittal}-e, the initial approach entailed the division of the mask into three regions of interest, according to the specific hernia or scar location for pre-operative and post-operative MRI respectively. The lowest region, called infra, encompasses the area from the bottom of the image to a point 3 cm below the largest hernia neck, where the axial plane was selected. The second region, called hernia-scar, was centered on the hernia neck (or scar) $\pm$ 3 cm. The upper and third region, called supra, extends from 3 cm above the hernia neck to the top of the image. This 3 cm-landmark has been commonly used in the literature \cite{hernandez-granadosEuropeanHerniaSociety2021}. This approach allows for intra-patient comparisons between pre-operative and post-operative stages, with a particular emphasis on the hernia region. The displacement was then averaged within these regions.

\figureDisplacementTimelineSagittal

\subsubsection{Statistical analysis}

\noindent The results are presented as an average of all participants ± one standard deviation noted $\sigma$, with their corresponding [minimum/maximum] range.

\noindent The differences between exercises (breathing, coughing, Valsalva), between linea alba regions (infra, hernia-scar, supra), and between pre-operative and post-operative stages were assessed for the maximum values of the variables. The maximum corresponds to the end of inhalation for breathing, cough peak for coughing, and onset plateau for Valsalva. For those comparisons, a paired t-test or Wilcoxon test was used according to the normality hypothesis validation of a Shapiro-Wilk test.

\vspace{0.1cm}

\noindent The effect of mesh placement (e.g IPOM, retro-rectus etc) on the different variables was evaluated. A one-way analysis of variance (ANOVA) or Kruskal-Wallis test was used according to the normality hypothesis validation of a Shapiro-Wilk test. If ANOVA yielded a significant result, Tukey’s post-hoc test was conducted for pairwise comparisons between the different mesh placements. If the Kruskal-Wallis test was significant, Dunn’s post-hoc test with Bonferroni correction was applied to identify specific mesh placement differences. Statistical significance was considered for \textit{p}-value $\leq$ $\PvaluethresholdPatients$. 

\noindent Additionally, the post-operative change ratio was defined as the difference between post-operative and pre-operative stages, with respect to the pre-operative stage $(Postop-Preop)/Preop$ * 100 $(\%)$.

\noindent Linear correlation between the maximum values of different variables was assessed by a Pearson correlation coefficients \textit{r} to evaluate the direction (positive or negative) and strength of correlations, with statistical significance threshold set to \textit{p} $\leq$ $\PvaluethresholdPatients$.

\vspace{0.1cm}

\noindent To further visualise the distribution of maximum values among participants before and after surgery, radar plots were employed to represent the spread of data, often obscured by the averages. These graphs aimed to highlight the variability in biomechanical changes induced by surgery across different participants, emphasising the individualised impact of the surgical intervention on abdominal wall dynamics.

\section{Results}

\subsection{Cohort description}

\noindent The cohort details are summarised in \autoref{tab:results_cohort_description_morpho_hernia_surgery}.

\begin{table}[H]
\renewcommand{\arraystretch}{2}
\centering
\begin{center}
\resizebox{\columnwidth}{!}{%
\begin{tabular}{ |c|c|c|c|c|c|c|c|c|c|c| } 
\hline
 & \multicolumn{4}{c|}{\textbf{Morphometry}} & \multicolumn{2}{c|}{\textbf{Hernia details}} & \multicolumn{4}{c|}{\textbf{Surgery details}} \\
\hline
\textbf{Participant} &
\makecell{\textbf{Sex} \\ \textbf{(M/F)}} & \textbf{Age $(yo)$} & \makecell{\textbf{Duration between} \\ \textbf{surgery and} \\ \textbf{post-op MRI} $(days)$} & \textbf{BMI $(\frac{kg}{m^2})$} &
\textbf{EHS classification} & \makecell{\textbf{Defect size} \\ (Height (cm) x width (cm))} &
\textbf{Surgery procedure} & \textbf{Mesh type} & \textbf{Mesh size ($cm^2$)} & \textbf{Mesh placement} \\
\hline
\hline
\textbf{1} & M & 73 & 362 & 29.1 & Incisional midline M2 & 3x4 & Laparotomy & Polyester* & 300 & IPOM+ \\
\hline
\textbf{2} & M & 71 & 103 & 38.9 & Incisional midline M2 & 3x3.8 & Laparotomy & Polyester* & 150 & IPOM+ \\
\hline
\textbf{3} & F & 51 & 125 & 30.8 & Incisional midline M3 & 3.9x6.7 & Laparotomy & Polyester & 500 & IPOM+ \\
\hline
\textbf{4} & F & 47 & 97 & 32.8 & Incisional midline M2 & 5.5x3.7 & Laparotomy & Self-gripping polyester* & 225 & Retro-rectus \\
\hline
\textbf{5} & M & 55 & 104 & 28.5 & Primary umbilical & 1.5x1.5 & Laparotomy & Polypropylene & 38 & Pre-peritoneal \\
\hline
\textbf{6} & M & 60 & 95 & 32.4 & Primary sus-umbilic & 3.5x6 & Laparotomy & Polypropylene & 38 & IPOM+ \\
\hline
\textbf{7} & F & 73 & 223 & 30.4 & Incisional midline M2 & 6.1x5 & Laparotomy & Polyester & 500 & Retro-rectus \\
\hline
\textbf{8} & M & 61 & 88 & 29.7 & Primary sus-umbilic & 2.5x4 & Laparotomy & Polypropylene & 64 & Pre-peritoneal \\
\hline
\textbf{9} & M & 28 & 76 & 30.7 & Incisional traumatic L4 & 10x8.2 & Laparotomy & Self-gripping polyester* & 225 & \makecell{Retro-muscular\\ Pre-peritoneal} \\
\hline
\textbf{10} & F & 75 & 125 & 27.7 & Incisional midline M4 & 3.2x4.2 & \makecell{Laparotomy and\\laparoscopy} & Polyester & 225 & IPOM+ \\
\hline
\textbf{11} & F & 42 & 111 & 28.0 & Incisional midline M3 & 3.7x3.1 & \makecell{Laparotomy and\\laparoscopy} & Polyester & 177 & IPOM+ \\
\hline
\end{tabular}
}
\end{center}
\captionsetup{justification=centering, format=plain}
\caption[Morphometric data of the patients]{Cohort description according to morphometric data (sex, age at pre-operative MRI, duration between surgery and post-operative MRI, BMI), hernia parameters (EHS classification \cite{muysomsClassificationPrimaryIncisional2009}, defect size) and surgery details (procedure type, as well as mesh type, size and placement). The "*" symbol next to the mesh type indicates that an absorbable film is included}
\label{tab:results_cohort_description_morpho_hernia_surgery}
\end{table}

\noindent The morphometric details of this study include sex, age at pre-operative MRI, the duration between surgery and post-operative MRI, and the BMI. The cohort was composed of 11 patients (6 females). The average age was $\meanagePatients$ $\pm$ $\stdagePatients$ years old (min = $\minagePatients$, max = $\maxagePatients$). The duration between surgery and post-operative MRI was $\meandurationpostopmri$ $\pm$ $\stddurationpostopmri$ days (min = $\mindurationpostopmri$, max = $\maxdurationpostopmri$ days). The BMI was $\meanBMIPatients$ $\pm$ $\stdBMIPatients$ $kg/m^2$ (min = $\minBMIPatients$, max = $\maxBMIPatients$ $kg/m^2$).

\vspace{0.1cm} 

\noindent The hernias were classified according to the EHS classification \cite{muysomsClassificationPrimaryIncisional2009}. Among the 11 patients, there were three primary midline umbilical hernias, four occurrences of incisional midline M2 hernia, two incisional midline M3 hernia, one incisional midline M4 hernia, and one incisional lateral L4 hernia. The average defect width was $\meanHerniaWidth$ $\pm$ $\stdHerniaWidth$ mm (min = $\minHerniaWidth$, max = $\maxHerniaWidth$ mm). The average defect height was $\meanHerniaHeight$ $\pm$ $\stdHerniaHeight$ mm (min = $\minHerniaHeight$, max = $\maxHerniaHeight$ mm). The width and height of defect were linearly correlated (\textit{r}=0.65, \textit{p} $\leq$ $\PvaluethresholdPatients$). 

\vspace{0.1cm} 

\noindent The surgery details are as follows. Most of the patients had a laparotomy (open surgery), while two of them both had laparoscopy to first place the mesh and then a laparotomy to fix it on the abdominal wall. All patients had surgery with defect closure with mesh reinforcement. The average mesh size was $\meanMeshSurface$ $\pm$ $\stdMeshSurface$ $cm^2$ (min = $\minMeshSurface$, max = $\maxMeshSurface$ $cm^2$).

\subsection{Pre-operative hernia sac description}

\noindent The hernia sac was fully segmented over time for 8 patients. For the remaining 3, the accuracy of the sac segmentation at each frame were considered insufficient to assess its dynamic evolution. The hernia sac description was performed for the pre-operative stage only, as there was no hernia post-operatively.

\noindent In axial plane, during the exercises, the maximum hernia sac area was $\meanHerniaSacpreBreathing$ $\pm$ $\stdHerniaSacpreBreathing$ $cm^2$, $\meanHerniaSacpreCoughing$ $\pm$ $\stdHerniaSacpreCoughing$ $cm^2$, $\meanHerniaSacpreValsalva$ $\pm$ $\stdHerniaSacpreValsalva$ $cm^2$ for breathing, coughing and Valsalva respectively. In relative terms, the area of the hernia sac increased by $\meanHerniaSacVariationpreBreathing$ $\pm$ $\stdHerniaSacVariationpreBreathing$$\%$, $\meanHerniaSacVariationpreCoughing$ $\pm$ $\stdHerniaSacVariationpreCoughing$$\%$ and $\meanHerniaSacVariationpreValsalva$ $\pm$ $\stdHerniaSacVariationpreValsalva$$\%$ during breathing, coughing and Valsalva respectively. 

\vspace{0.1cm}

\noindent In sagittal plane, during Valsalva the maximum hernia sac area was $\meanSagHerniaSacpreValsalva$ $\pm$ $\stdSagHerniaSacpreValsalva$ $cm^2$, and the corresponding relative increase $\meanSagHerniaSacVariationpreValsalva$ $\pm$ $\stdSagHerniaSacVariationpreValsalva$$\%$.

\noindent There was a strong correlation between axial and sagittal hernia sac areas during Valsalva (\textit{r}=0.97, \textit{p} $\leq$ $\PvaluethresholdPatients$). No correlation was found between the defect dimensions and the maximum hernia sac area, neither in axial plane nor in sagittal plane.

\subsection{Inter-recti distance and angle}

\noindent The inter-recti distance and angle are illustrated in \autoref{fig:Results_diastasis_angle_inter_vol_pre_post} together with the corresponding time changes before and after surgery. The corresponding maximum for each patient is displayed in the circular plots. The patient \#9 with the lateral hernia was excluded from the analysis due to the focus on the rectus abdominis region as measured by these metrics. These two variables were then assessed on 10 patients.

\noindent On average, the inter-recti distance was reduced by $\meanRAMDiffAllExercises$ $\pm$ $\stdRAMDiffAllExercises$ mm after the surgery, which corresponds to a reduction of $\meanRatioRAMAllExercises \pm \stdRatioRAMAllExercises\%$. The inter-recti angle was reduced by $\meanDefectAngleDiffAllExercises$ $\pm$ $\stdDefectAngleDiffAllExercises$$^{\circ}$ approximately. As illustrated in the circular plots \autoref{fig:Results_diastasis_angle_inter_vol_pre_post}-b, a large range of post-operative differences was observed among patients.

\figureInterRectiAngle

\noindent The maximum value of inter-recti distance and angle, averaged among the 10 patients, are reported in \autoref{tab:results_inter_recti_hernia_defect_angle}. The inter-recti distance was shorter after the surgery (\textit{p} = $\pvalRAMBreathing$ for breathing, \textit{p} $\leq$ $\PvaluethresholdPatients$ for coughing and Valsalva). As summarised in \autoref{tab:results_inter_recti_hernia_defect_angle}, the inter-recti angle was significantly smaller after the surgery for all the exercises (\textit{p} $\leq$ $\PvaluethresholdPatients$). 

\begin{table}[H]
\renewcommand{\arraystretch}{1.5}
\centering
\begin{center}
\begin{tabular}{ |c|c|c|c| } 
\hline
\textbf{Exercise} & \textbf{Pre-operative} & \textbf{Post-operative} & \textbf{\textit{p}-value} \\
\hline
\multicolumn{4}{|c|}{\textbf{Inter-recti distance $(mm)$}} \\
\hline
\textbf{Breathing} & $\meanRAMpreBreathing$ $\pm$ $\stdRAMpreBreathing$ & $\meanRAMpostBreathing$ $\pm$ $\stdRAMpostBreathing$ & $\pvalRAMBreathing$* \\
\hline
\textbf{Coughing} & $\meanRAMpreCoughing$ $\pm$ $\stdRAMpreCoughing$ & $\meanRAMpostCoughing$ $\pm$ $\stdRAMpostCoughing$ & $\pvalRAMCoughing$* \\
\hline
\textbf{Valsalva maneuver} & $\meanRAMpreValsalva$ $\pm$ $\stdRAMpreValsalva$ & $\meanRAMpostValsalva$ $\pm$ $\stdRAMpostValsalva$ & $\pvalRAMValsalva$* \\
\hline
\multicolumn{4}{|c|}{\textbf{Inter-recti angle $(^{\circ})$}} \\
\hline
\textbf{Breathing} & $\meanDefectAnglepreBreathing$ $\pm$ $\stdDefectAnglepreBreathing$ & $\meanDefectAnglepostBreathing$ $\pm$ $\stdDefectAnglepostBreathing$ & $\pvalDefectAngleBreathing$* \\
\hline
\textbf{Coughing} & $\meanDefectAnglepreCoughing$ $\pm$ $\stdDefectAnglepreCoughing$ & $\meanDefectAnglepostCoughing$ $\pm$ $\stdDefectAnglepostCoughing$ & $\pvalDefectAngleCoughing$* \\
\hline
\textbf{Valsalva maneuver} & $\meanDefectAnglepreValsalva$ $\pm$ $\stdDefectAnglepreValsalva$ & $\meanDefectAnglepostValsalva$ $\pm$ $\stdDefectAnglepostValsalva$ & $\pvalDefectAngleValsalva$* \\
\hline
\end{tabular}
\end{center}
\captionsetup{justification=centering, format=plain}
\caption[Pre-operative and post-operative inter-recti distance and angle]{Inter-recti distance and angle before and after surgery assessed at exercise peak (i.e., end of inhalation for breathing, cough peak for coughing, onset plateau for Valsalva) \\ Results are presented as mean values (n=10) $\pm$ one standard deviation $\sigma$, the star symbol (*) indicates a significant \textit{p}-value $\leq$ $\PvaluethresholdPatients$}
\label{tab:results_inter_recti_hernia_defect_angle}
\end{table}

\subsection{Visceral area}

\noindent The temporal evolution of visceral area before and after surgery is shown in \autoref{fig:Results_visceral_area_inter_vol_pre_post} for the three exercises in axial plane, and for Valsalva in sagittal plane. The corresponding maximum values for each patient are displayed in the circular plots. The sagittal visceral area during Valsalva followed a 3-phase pattern, i.e., increase, plateau and decrease.

\figureVisceralArea

\noindent The maximum values of visceral area, averaged among the 11 patients, are shown in \autoref{tab:results_visceral_area}. The post-operative increase in axial visceral area during breathing exercise was almost significant (\textit{p} = $\pvalVisceralAreaBreathing$).

\begin{table}[H]
\renewcommand{\arraystretch}{2}
\centering
\begin{center}
\begin{tabular}{ |c|c|c|c| } 
\hline
\textbf{Exercise} & \textbf{Pre-operative} & \textbf{Post-operative} & \textbf{\textit{p}-value} \\
\hline
\multicolumn{4}{|c|}{\textbf{Axial visceral area $(cm^2)$}} \\
\hline
\textbf{Breathing} & $\meanVisceralAreapreBreathing$ $\pm$ $\stdVisceralAreapreBreathing$ & $\meanVisceralAreapostBreathing$ $\pm$ $\stdVisceralAreapostBreathing$ & $\pvalVisceralAreaBreathing$ \\
\hline
\textbf{Coughing} & $\meanVisceralAreapreCoughing$ $\pm$ $\stdVisceralAreapreCoughing$ & $\meanVisceralAreapostCoughing$ $\pm$ $\stdVisceralAreapostCoughing$ & $\pvalVisceralAreaCoughing$ \\
\hline
\textbf{Valsalva maneuver} & $\meanVisceralAreapreValsalva$ $\pm$ $\stdVisceralAreapreValsalva$ & $\meanVisceralAreapostValsalva$ $\pm$ $\stdVisceralAreapostValsalva$ & $\pvalVisceralAreaValsalva$ \\
\hline
\multicolumn{4}{|c|}{\textbf{Sagittal visceral area $(cm^2)$}} \\
\hline
\textbf{Valsalva maneuver} & $\meanSagVisceralAreapreValsalva$ $\pm$ $\stdSagVisceralAreapreValsalva$ & $\meanSagVisceralAreapostValsalva$ $\pm$ $\stdSagVisceralAreapostValsalva$ & $\pvalSagVisceralAreaValsalva$ \\
\hline
\end{tabular}
\end{center}
\captionsetup{justification=centering, format=plain}
\caption[Pre-operative and post-operative visceral area]{Visceral area before and after surgery assessed at exercise peak (i.e., end of inhalation for breathing, cough peak for coughing, onset plateau for Valsalva) \\ Results are presented as mean values (n=11) $\pm$ one standard deviation $\sigma$, the star symbol (*) indicates a significant \textit{p}-value $\leq$ $\PvaluethresholdPatients$}
\label{tab:results_visceral_area}
\end{table}

\subsection{Axial displacement of abdominal muscles}

\noindent The axial displacements of both LM and RA before and after surgery are shown in \autoref{fig:Results_radial_disp_inter_vol_pre_post}. The circular plots (\autoref{fig:Results_radial_disp_inter_vol_pre_post}-b and \autoref{fig:Results_radial_disp_inter_vol_pre_post}-d) show the maximum displacement before and after the surgery for each patient, for LM and RA muscles respectively. These plots highlight a large spread of post-operative changes among patients, some having their mobility increased and some reduced.

\noindent The temporal evolution analysis during the exercise's execution indicated that both LM and RA muscles were moving outward during breathing, following a bell-shape pattern. The RA muscles moved more than LM. In addition, LM had a larger deviation corridor because some patients had the LM going outward, and some inward resulting in a negative displacement. This can be visualised in \autoref{fig:Results_radial_disp_inter_vol_pre_post}-b by looking at the patients outside and inside the circle of radius equals to zero.

\noindent During coughing, LM and RA had an opposite displacement. The LM muscles were going inward for almost all patients (except a very small displacement for patient \#11 visible \autoref{fig:Results_radial_disp_inter_vol_pre_post}-b). On the contrary, the RA muscles moved outward and the corresponding displacement magnitude was approximately twice that of the LM.

\noindent During Valsalva, LM and RA also had an opposite displacement, with a 3-phase pattern (increase-plateau-decrease). For all patients, the LM muscles were going inward whereas the RA muscles moved outward, again with a ratio of about two on the magnitudes.

\figureRadialDisp

\noindent \autoref{tab:results_muscle_displacements} shows the maximum displacement values for each exercise and each muscle group (LM or RA), averaged among the patients. During Valsalva, the post-operative displacement of LM was significantly larger (in magnitude) compared to the pre-operative one (\textit{p}=$\pvalRadialDispLMValsalva$).

\noindent The post-operative increased displacement of RA was almost significant during breathing (\textit{p}=$\pvalRadialDispRABreathing$).

\vspace{0.1cm}

\begin{table}[H]
\renewcommand{\arraystretch}{1.5}
\centering
\begin{center}
\begin{tabular}{ |c|c|c|c|c| } 
\hline
\textbf{Exercise} & \textbf{Muscle group} & \textbf{Pre-operative} & \textbf{Post-operative} & \textbf{\textit{p}-value} \\
\hline
\multicolumn{5}{|c|}{\textbf{Axial displacement $(mm)$}} \\
\hline
\multirow{2}*{\textbf{Breathing}} & \textcolor{red}{\textbf{LM}} & $\meanRadialDispLMpreBreathing$ $\pm$ $\stdRadialDispLMpreBreathing$ & $\meanRadialDispLMpostBreathing$ $\pm$ $\stdRadialDispLMpostBreathing$ & $\pvalRadialDispLMBreathing$ \\
& \textcolor{blue}{\textbf{RA}} & $\meanRadialDispRApreBreathing$ $\pm$ $\stdRadialDispRApreBreathing$ & $\meanRadialDispRApostBreathing$ $\pm$ $\stdRadialDispRApostBreathing$ & $\pvalRadialDispRABreathing$ \\
\hline
\multirow{2}*{\textbf{Coughing}} & \textcolor{red}{\textbf{LM}} & $\meanRadialDispLMpreCoughing$ $\pm$ $\stdRadialDispLMpreCoughing$ & $\meanRadialDispLMpostCoughing$ $\pm$ $\stdRadialDispLMpostCoughing$ & $\pvalRadialDispLMCoughing$ \\
& \textcolor{blue}{\textbf{RA}} & $\meanRadialDispRApreCoughing$ $\pm$ $\stdRadialDispRApreCoughing$ & $\meanRadialDispRApostCoughing$ $\pm$ $\stdRadialDispRApostCoughing$ & $\pvalRadialDispRACoughing$ \\
\hline
\multirow{2}*{\textbf{Valsalva}} & \textcolor{red}{\textbf{LM}} & $\meanRadialDispLMpreValsalva$ $\pm$ $\stdRadialDispLMpreValsalva$ & $\meanRadialDispLMpostValsalva$ $\pm$ $\stdRadialDispLMpostValsalva$ & $\pvalRadialDispLMValsalva$* \\
& \textcolor{blue}{\textbf{RA}} & $\meanRadialDispRApreValsalva$ $\pm$ $\stdRadialDispRApreValsalva$ & $\meanRadialDispRApostValsalva$ $\pm$ $\stdRadialDispRApostValsalva$ & $\pvalRadialDispRAValsalva$ \\
\hline
\end{tabular}
\end{center}
\captionsetup{justification=centering, format=plain}
\caption[Pre-operative and post-operative muscle displacement values]{Maximum displacement values of lateral muscles (LM) and rectus abdominis (RA) during breathing, coughing and Valsalva peak exercise \\ Results are presented as mean values (n=11) $\pm$ one standard deviation $\sigma$, the star symbol (*) indicates a significant \textit{p}-value $\leq$ $\PvaluethresholdPatients$}
\label{tab:results_muscle_displacements}
\end{table}

\noindent Mesh size impacted rectus muscles maximum post-operative displacement, with a significant negative correlation during breathing and coughing exercises (\textit{r}=$\pearsonRadialDispRApostMeshSizeBreathing$, \textit{p} = $\pvalRadialDispRApostMeshSizeBreathing$ $\leq$ $\PvaluethresholdPatients$ for breathing, \textit{r}=$\pearsonRadialDispRApostMeshSizeCoughing$, \textit{p} = $\pvalRadialDispRApostMeshSizeCoughing$ $\leq$ $\PvaluethresholdPatients$ for coughing). For this correlation study with mesh size, the patient \#9 with lateral hernia was also discarded (n=10).

\subsection{Shape of abdominal muscles}

\noindent \autoref{tab:results_muscle_strains} shows the maximum abdominal muscle length, thickness, circumferential and radial strains.

\noindent The maximum length of both LM and RA muscles was higher after surgery than before. For LM muscles, this increase was significant for coughing and Valsalva (\textit{p} $\leq$ $\PvaluethresholdPatients$), and almost significant for breathing (\textit{p} = 0.066). For RA muscles, the post-operative increase in length was significant for the three exercises (\textit{p} $\leq$ $\PvaluethresholdPatients$).

\vspace{0.1cm}

\noindent Regarding thickness, circumferential strain $\varepsilon_{\theta}$ and radial strain $\varepsilon_r$, no significant difference was observed between pre-operative and post-operative stages and so for both LM and RA muscles.

\vspace{0.1cm}

\noindent The post-operative maximum radial strains of RA $\varepsilon_r$ among the whole cohort were negatively correlated with the pre-operative defect widths during breathing (\textit{r}=-0.67, \textit{p} $\leq$ $\PvaluethresholdPatients$). Additionally, also for the breathing exercise, post-operative maximum radial strains of RA were found to be significantly impacted by the mesh placement in retro-rectus position compared to IPOM and pre-peritoneal placement (\textit{p} $\leq$ $\PvaluethresholdPatients$). Indeed, it was found to be significantly lower for patients who had a retro-rectus mesh placement. However, the defect widths were also significantly higher for patients who had a retro-rectus placement.

\begin{table}[H]
\renewcommand{\arraystretch}{1.2}
\centering
\begin{center}
\begin{tabular}{ |c|c|c|c|c| } 
\hline
\textbf{Exercise} & \textbf{Muscle group} & \textbf{Pre-operative} & \textbf{Post-operative} & \textbf{\textit{p}-value} \\
\hline
\multicolumn{5}{|c|}{\textbf{Length $(mm)$}} \\
\hline
\multirow{2}*{\textbf{Breathing}} & \textcolor{red}{\textbf{LM}} & $\meanLengthLMpreBreathing$ $\pm$ $\stdLengthLMpreBreathing$ & $\meanLengthLMpostBreathing$ $\pm$ $\stdLengthLMpostBreathing$ & 0.07 \\
& \textcolor{blue}{\textbf{RA}} & $\meanLengthRApreBreathing$ $\pm$ $\stdLengthRApreBreathing$ & $\meanLengthRApostBreathing$ $\pm$ $\stdLengthRApostBreathing$ & 0.01* \\
\hline
\multirow{2}*{\textbf{Coughing}} & \textcolor{red}{\textbf{LM}} & $\meanLengthLMpreCoughing$ $\pm$ $\stdLengthLMpreCoughing$ & $\meanLengthLMpostCoughing$ $\pm$ $\stdLengthLMpostCoughing$ & 0.02* \\
& \textcolor{blue}{\textbf{RA}} & $\meanLengthRApreCoughing$ $\pm$ $\stdLengthRApreCoughing$ & $\meanLengthRApostCoughing$ $\pm$ $\stdLengthRApostCoughing$ & 0.04* \\
\hline
\multirow{2}*{\textbf{Valsalva}} & \textcolor{red}{\textbf{LM}} & $\meanLengthLMpreValsalva$ $\pm$ $\stdLengthLMpreValsalva$ & $\meanLengthLMpostValsalva$ $\pm$ $\stdLengthLMpostValsalva$ & 0.03* \\
& \textcolor{blue}{\textbf{RA}} & $\meanLengthRApreValsalva$ $\pm$ $\stdLengthRApreValsalva$ & $\meanLengthRApostValsalva$ $\pm$ $\stdLengthRApostValsalva$ & 0.03* \\
\hline
\multicolumn{5}{|c|}{\textbf{Thickness $(mm)$}} \\
\hline
\multirow{2}*{\textbf{Breathing}} & \textcolor{red}{\textbf{LM}} & $\meanThicknessLMpreBreathing$ $\pm$ $\stdThicknessLMpreBreathing$ & $\meanThicknessLMpostBreathing$ $\pm$ $\stdThicknessLMpostBreathing$ & 0.54 \\
& \textcolor{blue}{\textbf{RA}} & $\meanThicknessRApreBreathing$ $\pm$ $\stdThicknessRApreBreathing$ & $\meanThicknessRApostBreathing$ $\pm$ $\stdThicknessRApostBreathing$ & 0.83 \\
\hline
\multirow{2}*{\textbf{Coughing}} & \textcolor{red}{\textbf{LM}} & $\meanThicknessLMpreCoughing$ $\pm$ $\stdThicknessLMpreCoughing$ & $\meanThicknessLMpostCoughing$ $\pm$ $\stdThicknessLMpostCoughing$ & 0.52 \\
& \textcolor{blue}{\textbf{RA}} & $\meanThicknessRApreCoughing$ $\pm$ $\stdThicknessRApreCoughing$ & $\meanThicknessRApostCoughing$ $\pm$ $\stdThicknessRApostCoughing$ & 0.49 \\
\hline
\multirow{2}*{\textbf{Valsalva}} & \textcolor{red}{\textbf{LM}} & $\meanThicknessLMpreValsalva$ $\pm$ $\stdThicknessLMpreValsalva$ & $\meanThicknessLMpostValsalva$ $\pm$ $\stdThicknessLMpostValsalva$ & 0.86 \\
& \textcolor{blue}{\textbf{RA}} & $\meanThicknessRApreValsalva$ $\pm$ $\stdThicknessRApreValsalva$ & $\meanThicknessRApostValsalva$ $\pm$ $\stdThicknessRApostValsalva$ & 0.41 \\
\hline
\multicolumn{5}{|c|}{\textbf{Circumferential strain $\varepsilon_{\theta}$ $(\%)$}} \\
\hline
\multirow{2}*{\textbf{Breathing}} & \textcolor{red}{\textbf{LM}} & $\meanCircStrainLMpreBreathing$ $\pm$ $\stdCircStrainLMpreBreathing$ & $\meanCircStrainLMpostBreathing$ $\pm$ $\stdCircStrainLMpostBreathing$ & 0.15 \\
& \textcolor{blue}{\textbf{RA}} & $\meanCircStrainRApreBreathing$ $\pm$ $\stdCircStrainRApreBreathing$ & $\meanCircStrainRApostBreathing$ $\pm$ $\stdCircStrainRApostBreathing$ & 0.24 \\
\hline
\multirow{2}*{\textbf{Coughing}} & \textcolor{red}{\textbf{LM}} & $\meanCircStrainLMpreCoughing$ $\pm$ $\stdCircStrainLMpreCoughing$ & $\meanCircStrainLMpostCoughing$ $\pm$ $\stdCircStrainLMpostCoughing$ & 0.29 \\
& \textcolor{blue}{\textbf{RA}} & $\meanCircStrainRApreCoughing$ $\pm$ $\stdCircStrainRApreCoughing$ & $\meanCircStrainRApostCoughing$ $\pm$ $\stdCircStrainRApostCoughing$ & 0.28 \\
\hline
\multirow{2}*{\textbf{Valsalva}} & \textcolor{red}{\textbf{LM}} & $\meanCircStrainLMpreValsalva$ $\pm$ $\stdCircStrainLMpreValsalva$ & $\meanCircStrainLMpostValsalva$ $\pm$ $\stdCircStrainLMpostValsalva$ & 0.35 \\
& \textcolor{blue}{\textbf{RA}} & $\meanCircStrainRApreValsalva$ $\pm$ $\stdCircStrainRApreValsalva$ & $\meanCircStrainRApostValsalva$ $\pm$ $\stdCircStrainRApostValsalva$ & 0.78 \\
\hline
\multicolumn{5}{|c|}{\textbf{Radial strain $\varepsilon_{r}$ $(\%)$}} \\
\hline
\multirow{2}*{\textbf{Breathing}} & \textcolor{red}{\textbf{LM}} & $\meanRadStrainLMpreBreathing$ $\pm$ $\stdRadStrainLMpreBreathing$ & $\meanRadStrainLMpostBreathing$ $\pm$ $\stdRadStrainLMpostBreathing$ & 0.43 \\
& \textcolor{blue}{\textbf{RA}} & $\meanRadStrainRApreBreathing$ $\pm$ $\stdRadStrainRApreBreathing$ & $\meanRadStrainRApostBreathing$ $\pm$ $\stdRadStrainRApostBreathing$ & 0.55 \\
\hline
\multirow{2}*{\textbf{Coughing}} & \textcolor{red}{\textbf{LM}} & $\meanRadStrainLMpreCoughing$ $\pm$ $\stdRadStrainLMpreCoughing$ & $\meanRadStrainLMpostCoughing$ $\pm$ $\stdRadStrainLMpostCoughing$ & 0.76 \\
& \textcolor{blue}{\textbf{RA}} & $\meanRadStrainRApreCoughing$ $\pm$ $\stdRadStrainRApreCoughing$ & $\meanRadStrainRApostCoughing$ $\pm$ $\stdRadStrainRApostCoughing$ & 0.08 \\
\hline
\multirow{2}*{\textbf{Valsalva}} & \textcolor{red}{\textbf{LM}} & $\meanRadStrainLMpreValsalva$ $\pm$ $\stdRadStrainLMpreValsalva$ & $\meanRadStrainLMpostValsalva$ $\pm$ $\stdRadStrainLMpostValsalva$ & 0.21 \\
& \textcolor{blue}{\textbf{RA}} & $\meanRadStrainRApreValsalva$ $\pm$ $\stdRadStrainRApreValsalva$ & $\meanRadStrainRApostValsalva$ $\pm$ $\stdRadStrainRApostValsalva$ & 0.58 \\
\hline
\end{tabular}
\end{center}
\captionsetup{justification=centering, format=plain}
\caption[Pre-operative and post-operative muscle strains]{Maximum thickness, length, radial and circumferential strains of lateral muscles (LM) and rectus abdominis (RA) during breathing, coughing and Valsalva peak exercise \\ Results are presented as mean values (n=11) $\pm$ one standard deviation $\sigma$, the star symbol (*) indicates a significant \textit{p}-value $\leq$ $\PvaluethresholdPatients$}
\label{tab:results_muscle_strains}
\end{table}

\subsection{Sagittal displacement of linea alba} \label{subsection_results_sag_disp}

\noindent The displacement of the three regions of linea alba (infra, hernia-scar, supra) before and after surgery during Valsalva maneuver is shown in \autoref{fig:Results_sagittal_disp_inter_vol_pre_post}. The average values for the whole cohort are displayed in \autoref{fig:Results_sagittal_disp_inter_vol_pre_post}-a. For visualisation purpose, the corridor response is plotted for each region in \autoref{fig:Results_sagittal_disp_inter_vol_pre_post}-c.

\noindent The linea alba displacement followed a 3-phase pattern (increase-plateau-decrease). In both pre-operative and post-operative stages, the displacement of the hernia-scar region was significantly larger than in the infra and supra regions (\textit{p} $\leq$ $\PvaluethresholdPatients$).

\noindent The maximum displacement of each linea alba region before and after surgery is displayed in \autoref{fig:Results_sagittal_disp_inter_vol_pre_post}-b for the 11 patients. As for axial displacement of LM and RA muscles, these circular plots highlighted the variability of post-operative changes among patients.

\noindent The post-operative displacement of the linea alba was lower only for the patients \#2 and \#8. These two patients are males, tended to have higher BMI, visceral area and abdominal muscle length. 

\figureLineaAlbaDisp

\noindent The maximum values of post-operative change ratio for each linea alba region and averaged among patients are summarised in \autoref{tab:results_LB_displacements_change}. The hernia-scar region underwent the largest post-operative change in displacement (in magnitude, either positive or negative), compared to the supra region which underwent the lowest (\textit{p} = $\pvalRatioSagDispHerniaSupraValsalva$).

\begin{table}[H]
\centering
\begin{tabular}{|c|c|c|}
\hline
\textbf{Region} & \makecell{\textbf{Post-operative change ratio $(\%)$} \\ ($Postop-Preop/Preop$ * 100)} & \textbf{\textit{p}-value of the paired t-test} \\ \hline
\textcolor{violet}{\textbf{Infra}} & $\meanRatioSagDispInfraValsalva$ $\pm$ $\stdRatioSagDispInfraValsalva$ & \multirow{2}{*}{\(\left.\rule{0mm}{4mm}\right\}\) \textit{p} = $\pvalRatioSagDispInfraHerniaValsalva$} \hspace{2cm} \multirow{3}{*}{\(\left.\rule{0mm}{6mm}\right\}\) \textit{p} = $\pvalRatioSagDispInfraSupraValsalva$} \\ \cline{1-2}
\textcolor{magenta}{\textbf{Hernia/scar}} & $\meanRatioSagDispHerniaValsalva$ $\pm$ $\stdRatioSagDispHerniaValsalva$ & \multirow{2}{*}{\(\left.\rule{0mm}{4mm}\right\}\) \textit{p} = $\pvalRatioSagDispHerniaSupraValsalva$} \\ \cline{1-2}
\textcolor{orange}{\textbf{Supra}} & $\meanRatioSagDispSupraValsalva$ $\pm$ $\stdRatioSagDispSupraValsalva$ & \\ \hline
\end{tabular}
\captionsetup{justification=centering, format=plain}
\caption[Post-operative change ratio of displacement of the linea alba]{Post-operative change ratio of the displacements depending on the linea alba region, assessed at Valsalva peak (i.e., onset plateau for Valsalva) \\ Results are presented as mean values (n=11) $\pm$ one standard deviation $\sigma$}
\label{tab:results_LB_displacements_change}
\end{table}
\section{Discussion}

\noindent This study is the first to evaluate both anatomical and functional changes in the abdominal wall before and after abdominal hernia surgery using dynamic MRI in axial and sagittal views. Pre-operative hernia sac as well as both pre-operative and post-operative visceral area, abdominal muscles and linea alba were visualised \textit{in vivo} in real-time thereby offering a more comprehensive understanding of post-surgical dynamic changes.

\vspace{0.2cm}

\noindent Hernia sac area changes were quantified during physical exercise.
The hernia sac area largely increased at coughing and Valsalva peaks ($\meanHerniaSacVariationpreCoughing$ $\pm$ $\stdHerniaSacVariationpreCoughing$$\%$ and $\meanHerniaSacVariationpreValsalva$ $\pm$ $\stdHerniaSacVariationpreValsalva$$\%$ respectively) in the axial plane. In the sagittal plane, the hernia sac area change during Valsalva ($\meanSagHerniaSacVariationpreValsalva$ $\pm$ $\stdSagHerniaSacVariationpreValsalva$$\%$) was quite similar to the one quantified in the axial plane. These results are supportive of CT scan-based results reported by Bellio \textit{et al.} \cite{bellioPreoperativeAbdominalComputed2019}. They actually found a 23$\%$ increase in defect area during the Valsalva contraction and an 82$\%$ increase in hernia sac volume.

\noindent In healthy subjects, coughing and Valsalva maneuver induce similar changes in intra-abdominal pressure \cite{soucasseBetterUnderstandingDaily2022}. Therefore, we could expect a similar increase in hernia sac area during coughing and the Valsalva maneuver. The difference between these two exercises may be due to the fact that the Valsalva maneuver was systematically performed after the breathing and coughing sessions. It could then occur that the hernia sac was remaining in an extended position after coughing thereby minimising the hernia sac changes during Valsalva. In addition, the variation of hernia sac area was very patient-specific thereby accounting for the very large standard variation.

\noindent In both axial and sagittal planes, there was no correlation between the defect size (respectively width and height) and the hernia sac area. This result indicates that the largest defect is not necessarily linked to the largest hernia sac area.

\vspace{0.2cm}

\noindent One of the contributions of this study is the quantification of the inter-recti distance and angle during various exercises, before and after the surgery. Systematic closure of the defect is not accompanied by a complete reduction in inter-recti distance, as illustrated by the $\meanRatioRAMAllExercises \pm \stdRatioRAMAllExercises\%$ post-operative reduction. Only patient \#8 had a post-operative inter-recti distance of zero, as the two rectus abdominis sheaths were brought completely together during suturing, leaving no space between them. However, a residual rectus diastasis, corresponding to an inter-recti distance of 20mm or more \cite{cavalliPrevalenceRiskFactors2021}, has been recognised as a significant risk factor for recurrence \cite{bittnerUpdateGuidelinesLaparoscopic2019,boothDiastasisRectiAssociated2023}. It would be interesting to examine the recurrence within this cohort and whether there is an association with the corresponding inter-recti distance. 

\noindent A significant decrease has been observed in the inter-recti angle in axial MRI views, which takes into account the patient's antero-poster depth. The combined effects of the surgical closure and the enhanced muscle mobility may play an important role in the reduction of the inter-recti angle post-operatively. 

\vspace{0.2cm}

\noindent The visceral area, quantified in axial and sagittal planes, showed a slight increase post-operatively during breathing in the axial plane (\textit{p} = $\pvalVisceralAreaBreathing$), likely due to hernia sac content being repositioned into the abdomen during defect closure. No differences were observed during exercises involving muscular contraction (e.g., cough, Valsalva) in either plane. While this metric provided limited insights into anatomical and functional changes in this cohort, it may hold relevance for recurrence studies or specific hernia types, such as giant hernias, where significant external abdominal content could markedly impact surgical outcomes when returned to the abdominal cavity \cite{sabbaghPeritonealVolumePredictive2011}.

\vspace{0.2cm}

\noindent Our results detailed the abdominal muscles motion over time before and after hernia surgery. In the axial plane, the displacement of rectus abdominis and lateral muscles followed a pattern which was specific to each exercise. At both pre-operative and post-operative stages, rectus abdominis muscles were more involved than lateral muscles during breathing. This result was also observed in healthy subjects using a similar dynamic MRI approach \cite{jourdanDynamicMRIQuantificationAbdominal2022}. During exercises such as coughing and Valsalva, rectus abdominis and lateral muscles had a mirrored pattern. Lateral muscles displayed a negative displacement (inward direction) whereas rectus muscles were pushed outward.

\noindent Of interest, our results also indicated differences in abdominal muscles displacement between pre-operative and post-operative stages. After surgery, a more pronounced inward displacement of lateral muscles during Valsalva was observed (\textit{p}= $\pvalRadialDispLMValsalva$). These changes might account for an optimised involvement of these muscles during physiological activities as a result of the surgery. They could illustrate an improved biomechanical function which could be crucial for patients’ post-operative recovery and long-term outcomes. 

\noindent The maximum post-operative displacement of the rectus abdominis muscles during breathing and coughing was negatively correlated with the mesh size. This could suggest that the bigger is the mesh, the more restricted is the displacement of the midline. These findings are supportive of previous studies showing that repaired abdominal walls tend to be more rigid \cite{dubayMeshIncisionalHerniorrhaphy2006,leruyetDifferencesBiomechanicsAbdominal2020} with reduced mobility compared to a healthy or herniated abdominal wall \cite{jungeElasticityAnteriorAbdominal2001,todrosComputationalModelingAbdominal2018,heNumericalMethodGuiding2020,hernandezMechanicalHistologicalCharacterization2011}. 
Le Ruyet \textit{et al.} \cite{leruyetDifferencesBiomechanicsAbdominal2020} demonstrated, using image correlation and known intra-abdominal pressures, that abdominal wall stiffness increases with greater mesh overlap. This is consistent with current surgical guidelines recommending larger mesh sizes with increased overlap relative to the defect size \cite{henriksenGuidelinesTreatmentUmbilical2020}.

\vspace{0.2cm}

\noindent This study also compared length and thickness dynamic changes of abdominal muscles before and after surgery. The corresponding differences indicated a significant lengthening after the surgical process which might be due to the surgery itself. In fact, the surgical closing of the hernia neck is related to a lengthening of the muscles as the surgeon intends to bring the defect sides close together. Of interest, the lengthening did not change the muscles deformability as assessed by the circumferential $\varepsilon_{\theta}$ and radial strains $\varepsilon_r$. This result may suggest that the surgery has changed the anatomy of the abdominal muscles but not their functional behaviour.

\noindent A negative correlation was observed between the post-operative rectus abdominis radial strain $\varepsilon_r$ and defect width during breathing, suggesting that larger defects are associated with less passive distension of the rectus abdominis after surgery. This finding aligns with the negative correlation we identified between mesh size and post-operative rectus abdominis displacement. Similar observations were reported by García Moriana \textit{et al.}, Gunnarsson \textit{et al.}, and Strigard \textit{et al.}, who found that larger defect widths or rectus diastasis correlated with reduced pre-operative rectus abdominis strength \cite{garciamorianaEvaluationRectusAbdominis2023,gunnarssonCorrelationAbdominalRectus2015,strigardGiantVentralHernia2016}. However, our results differ as they pertain to breathing—a passive exercise not requiring active muscle contraction—rather than muscle strength, limiting direct comparisons with these earlier studies. Notably, in our cohort, defect size, which determines the surgical approach, appears to influence the post-operative biomechanics of the rectus abdominis more than its pre-operative function.
Additionally, for patient's subgroup with mesh placement in retro-rectus position, the post-operative maximum radial thickness change of RA during breathing exercise was found to be significantly lower and defect widths higher. That said, interpreting these results remains complex, as this patient's subgroup also had significantly different pre-operative rectus thickness at rest (i.e. without contraction). As a result, the observed effect may be a combined influence of both pre-existing anatomical differences and surgical intervention, rather than an isolated effect of mesh placement alone. Due to these interdependencies within a small cohort, further studies with larger sample sizes are needed to disentangle the specific role of each factor.

\vspace{0.2cm}

\noindent This study also focused on the linea alba motion before and after hernia surgery, measured in the sagittal plane. The difference in displacement between pre and post-operative stages was not statistically significant. Our results showed a large range of post-operative changes among the patients, with either unchanged, increased or decreased mobility.

\noindent The highest post-operative change was observed in the hernia-scar region; i.e., the region which underwent the surgery; confirming the impact of the surgery on the dynamic behaviour of the abdominal wall. 

\noindent A careful look at the results indicates that the two patients with a reduced displacement of the linea alba post-surgery were males with a high BMI, visceral area and abdominal muscle length. Of interest, multiple studies have shown that these characteristics were associated with a higher risk of hernia recurrence \cite{oweiImpactBodyMass2017,vansilfhoutRecurrentIncisionalHernia2021,sorensenSmokingRiskFactor2005,lehuunhoIncidencePreventionVentral2012}. These findings raise the question of whether patients with reduced linea alba displacement post-surgery may eventually develop a recurrent hernia. This highlights the potential for post-operative mobility metrics to serve as predictive indicators for recurrence and warrants further investigation in long-term follow-up studies.

\vspace{0.6cm}

\noindent Several limitations should be acknowledged in the present study. While this study introduces a novel method for assessing functional outcomes pre- and post-operatively, the current sample size, large variability between subjects, and lack of long-term follow-up limit the generalisation of our findings and the immediate clinical recommendations that can be made. With a larger and more diverse cohort, as well as longitudinal data, further conclusions could be drawn to refine surgical decision-making. Additionally, we quantified significant changes for some metrics after the surgery and the long-term durability of these changes remains to be studied. The recovery from hernia surgery involves complex interactions between patient anatomy, hernia characteristics, and surgical techniques, which may not manifest uniformly across metrics. The study focuses on anatomical and biomechanical changes, but incorporating functional outcomes (e.g., pain, quality of life, physical activity levels) would provide a more comprehensive assessment of these interactions and of the surgical impact. Exercises were performed in a uniform way and this non-randomised aspect might have impacted the results even if the comparative analysis remains valid. In future studies, hernia sac retractation might be performed before each exercise for the sake of standardisation. 

\noindent Our study was conducted in single axial and sagittal planes. The axial plane was chosen where the hernia neck was the largest and the sagittal plane was placed in the middle on the hernia neck, often within the linea alba. In that configuration, the motion of muscles in the sagittal plane could not be assessed. A full 3D dynamic analysis could provide a more complete understanding of abdominal wall motion, keeping in mind that the time resolution would be largely compromised.

\noindent In addition, there is a need to further develop the automation of these semi-automated tools to facilitate their integration and use in a clinical context. However, it is important to acknowledge the economic limitations of dynamic MRI. Compared to CT, MRI is more expensive and less widely available, which may limit its routine clinical use for evaluating abdominal hernias. Nevertheless, MRI offers superior soft tissue contrast and avoids ionizing radiation, making it particularly valuable in specific cases where detailed functional and anatomical assessment is required. Future ongoing research is exploring ways to optimize imaging protocols, reduce costs, and improve accessibility to ensure broader clinical applicability.

\vspace{0.2cm}

\noindent Patient-specific factors such as BMI have been briefly addressed in this study. Morphometric characteristics such as BMI, age, and comorbidities (e.g., diabetes, smoking) could potentially influence post-operative biomechanics, and future studies with larger cohorts will be necessary to better assess their potential impact on surgical outcomes.

\noindent Mesh placement was also briefly introduced and its impact on post-operative outcomes. However, due to the small cohort of this study and the amount of interdependent factors, it is challenging to draw definitive conclusions based on categorical subgroup analyses (e.g., according to mesh placement or type). Extending these findings to larger cohorts and exploring the impact of mesh placement and surgery technique on long-term surgery outcomes may be of great interest. It would be also of interest to examine mesh position on MRI scans. 

\section{Conclusion}

\noindent This study provides a novel and comprehensive assessment of abdominal wall biomechanics before and after hernia surgery using dynamic MRI in both axial and sagittal planes. The findings disclose significant differences after surgery related to abdominal muscles and linea alba displacement, and an influence of defect and mesh size on post-operative biomechanics. It should be noted that there is a large inter-patient variability, which would need to be further investigated. These variables could be considered as quantitative indices of surgical outcome to be confirmed in case of recurrence. Furthermore, to facilitate their integration and use in a clinical context, the automation of these semi-automated tools needs to be further developed. If these results are confirmed in a larger number of subjects, this quantitative approach could be of great interest in clinical practice for a more comprehensive evaluation of hernia surgery.

\section*{Data availability statement}

\noindent The data presented in this study is stored in a public repository and can be made available on a reasonable request from the corresponding author.

\noindent The DOI associated with the dataset is: \href{https://doi.org/10.57745/KTM2OA}{10.57745/KTM2OA}.

\section{Appendix}

\subsection{MRI parameters} \label{appendix:appendix_MRI_parameters}

\noindent The duration between the surgery and the post-operative MRI, as well as acquisition parameters such as field of view, temporal and spatial resolution for both axial and sagittal MR sequences are reported in \autoref{tab:results_acquisition_parameters}. The acquisition parameters were the same across the exercises.

\begin{table}[H]
\renewcommand{\arraystretch}{1.5}
\centering
\begin{center}
\begin{tabular}{ |c|c|c| } 
\hline
\textbf{Data} & \textbf{Average $\pm$ $\sigma$} & \textbf{Range [min/max]} \\
\hline
\multicolumn{3}{|c|}{\textbf{Axial MRI}} \\
\hline
\textbf{Field of view $(mm)$} & $\meanFoVaxial$ $\pm$ $\stdFoVaxial$ & [$\minFoVaxial$ / $\maxFoVaxial$] \\
\hline
\textbf{Temporal resolution $(ms)$} & $\meanTRaxial$ $\pm$ $\stdTRaxial$ & [$\minTRaxial$ / $\maxTRaxial$] \\
\hline
\textbf{Spatial resolution $(mm)$} & $\meanpXaxial$ $\pm$ $\stdpXaxial$ & [$\minpXaxial$ / $\maxpXaxial$] \\
\hline
\multicolumn{3}{|c|}{\textbf{Sagittal MRI}} \\
\hline
\textbf{Field of view $(mm)$} & $\meanFoVsagittal$ $\pm$ $\stdFoVsagittal$ & [$\minFoVsagittal$ / $\maxFoVsagittal$] \\
\hline
\textbf{Temporal resolution $(ms)$} & $\meanTRsagittal$ $\pm$ $\stdTRsagittal$ & [$\minTRsagittal$ / $\maxTRsagittal$] \\
\hline
\textbf{Spatial resolution $(mm)$} & $\meanpXsagittal$ $\pm$ $\stdpXsagittal$ & [$\minpXsagittal$ / $\maxpXsagittal$] \\
\hline
\end{tabular}
\end{center}
\captionsetup{justification=centering, format=plain}
\caption[MRI parameters]{Acquisition parameters of pre-operative MRI \\ Results are presented as mean values $\pm$ one standard deviation $\sigma$}
\label{tab:results_acquisition_parameters}
\end{table}

\noindent The average duration of sequences is 41 $\pm$ 2s, 23 $\pm$ 3s and 54 $\pm$ 6s for breathing, coughing and Valsalva respectively.

\section*{Citation Diversity Statement}

\noindent Recent work in several fields of science has identified a bias in citation practices such that papers from women and other minority scholars are under-cited relative to the number of such papers in the field \cite{mitchell2013gendered,dion2018gendered,caplar2017quantitative, maliniak2013gender, Dworkin2020.01.03.894378, bertolero2021racial, wang2021gendered, chatterjee2021gender, fulvio2021imbalance}. Here we sought to proactively consider choosing references that reflect the diversity of the field in thought, form of contribution, gender, race, ethnicity, and other factors.

\vspace{0.3cm}

\noindent First, we obtained the predicted gender of the first and last author of each reference by using databases that store the probability of a first name being carried by a woman \cite{Dworkin2020.01.03.894378,zhou_dale_2020_3672110}. By this measure (and excluding self-citations to the first and last authors of our current paper), our references contain 7.82\% woman(first)/woman(last), 14.57\% man/woman, 19.84\% woman/man, and 57.77\% man/man. This method is limited in that a) names, pronouns, and social media profiles used to construct the databases may not, in every case, be indicative of gender identity and b) it cannot account for intersex, non-binary, or transgender people.

\vspace{0.3cm}

\noindent Second, we obtained predicted racial/ethnic category of the first and last author of each reference by databases that store the probability of a first and last name being carried by an author of color \cite{ambekar2009name, sood2018predicting}. By this measure (and excluding self-citations), our references contain 17.93\% author of color (first)/author of color(last), 18.59\% white author/author of color, 16.50\% author of color/white author, and 46.98\% white author/white author. This method is limited in that a) names and Florida Voter Data to make the predictions may not be indicative of racial/ethnic identity, and b) it cannot account for Indigenous and mixed-race authors, or those who may face differential biases due to the ambiguous racialization or ethnicization of their names. We look forward to future work that could help us to better understand how to support equitable practices in science.

\section*{Funding}

\noindent This research was supported by Université Gustave Eiffel.
\section*{Acknowledgments}

\noindent The authors would like to thank the collaborators of the CRMBM/CEMEREM UMR CNRS 7339 for their contribution to this research, especially to Stanislas Rapacchi, Pierre Daude, and Constance Michel for their insightful help and comments on the data processing. The authors would also like to thank Kathia Chaumoitre, Mathieu Di Bisceglie, Rym Djouri as well as all the radiology technicians of the North Hospital of Marseille, who greatly facilitated the MRI acquisitions for the purpose of this study.

\vspace{0.5cm}

\noindent This version of the article has been accepted for publication, after peer review and is subject to Springer Nature’s \href{https://www.springernature.com/gp/open-research/policies/accepted-manuscript-terms}{AM terms of use}, but is not the Version of Record and does not reflect post-acceptance improvements, or any corrections. The Version of Record is available online at: \href{https://doi.org/10.1007/s10029-025-03337-4}{10.1007/s10029-025-03337-4}
\section*{Statements and Declarations}

\noindent \textbf{Conflict of interest}: The authors declare that they have no conflict of interest.

\noindent \textbf{Ethical approval}:  The Local Ethics Committee approved this study.

\noindent \textbf{Consent to participate}: All participants in this study provided informed consent.
\section*{Author contributions}

\noindent \textbf{VJ}: Conceptualisation, Data curation, Formal analysis, Funding acquisition, Investigation, Methodology, Software, Visualization, Writing - original draft

\noindent \textbf{DB}, \textbf{CM} and \textbf{TB}: Conceptualisation, Funding acquisition, Methodology, Project administration, Resources, Writing - review \& editing

\noindent \textbf{AAEA}: Conceptualisation, Funding acquisition, Resources, Writing - review \& editing

\newpage
\bibliography{clean_and_diversity_merged.bib}

\end{document}